\documentclass[journal]{IEEEtran}
\IEEEoverridecommandlockouts
\usepackage{amsmath}
\usepackage{amsfonts}
\usepackage{amssymb}
\usepackage[english]{babel}
\usepackage{amsthm}
\newtheorem{remark}{Remark}

\newtheorem{lemma}{Lemma}
\newtheorem{theorem}{Theorem}
\usepackage{esint}
\usepackage{algorithm,algorithmicx}
\usepackage{algpseudocode}
\usepackage{array}
\usepackage[caption=false,font=normalsize,labelfont=sf,textfont=sf]{subfig}
\usepackage{textcomp}
\usepackage{stfloats}
\usepackage{multirow}
\usepackage{booktabs}
\usepackage{etoolbox}
\usepackage{geometry}
\usepackage{pstricks}
\usepackage{svg}

\usepackage{url}
\usepackage{verbatim}
\newtheorem{definition}{Definition}
\usepackage{cite}
\def\BibTeX{{\rm B\kern-.05em{\sc i\kern-.025em b}\kern-.08em
    T\kern-.1667em\lower.7ex\hbox{E}\kern-.125emX}}
\usepackage{subfloat}
\usepackage{tikz}
\usetikzlibrary{graphs, arrows, positioning, quotes, shapes.geometric, arrows.meta}

\ifCLASSINFOpdf
\else
\fi

\begin{document}

\title{ Frequency-Space Channel Estimation and Spatial Equalization in Wideband Fluid Antenna System}

\author{
Xuehui~Dong,~\IEEEmembership{Student Member,~IEEE,}  
Kai~Wan,~\IEEEmembership{Member,~IEEE,} 
Shuangyang~Li,~\IEEEmembership{Member,~IEEE,}
Robert~Caiming~Qiu,~\IEEEmembership{Fellow,~IEEE},
and~Giuseppe Caire,~\IEEEmembership{Fellow,~IEEE}
\thanks{X.~Dong, K.~Wan, and R.~C.~Qiu are  with the School of Electronic Information and Communications, 
Huazhong University of Science and Technology, 430074  Wuhan, China,  (e-mail: \{xuehuidong, kai\_wan, caiming\}@hust.edu.cn).}
\thanks{
 S.~Li and G.~Caire are with the Electrical Engineering and Computer
Science Department, Technische Universit\"at Berlin, 10587 Berlin, Germany
(e-mail:  \{shuangyang, caire\}@tu-berlin.de).}
}


\maketitle

\begin{abstract}
The Fluid Antenna System (FAS) overcomes the spatial degree-of-freedom limitations of conventional static antenna arrays in wireless communications.This capability critically depends on acquiring full Channel State Information across all accessible ports. Existing studies focus exclusively on narrowband FAS, performing channel estimation solely in the spatial domain.
This work proposes a channel estimation and spatial equalization framework for wideband FAS, revealing for the first time an inherent group-sparse structure in aperture-limited FAS channels.
First, we establish a group-sparse recovery framework for space-frequency characteristics in FAS, formally characterizing leakage-induced sparsity degradation from limited aperture and bandwidth as a structured group-sparsity problem. By deriving dictionary-adapted group restricted isometry property, we prove tight recovery bounds for a convex $\ell_1/\ell_2$-mixed norm optimization formulation that preserves leakage-aware sparsity patterns. Second, we develop a descending correlation group orthogonal matching pursuit algorithm that systematically relaxes leakage constraints to reduce subcoherence. This approach enables FSC recovery with accelerated convergence and superior performance compared to conventional compressive sensing methods like OMP or GOMP. Third, we formulate spatial equalization as a mixed-integer linear programming problem, complement this with a greedy algorithm maintaining near-optimal performance. Simulation results demonstrate the proposed channel estimation algorithm effectively resolves energy misallocation and enables recovery of weak details, achieving superior recovery accuracy and convergence rate. The SE framework suppresses deep fading phenomena and largely reduces time consumption overhead while maintaining equivalent link reliability.
\end{abstract}

\begin{IEEEkeywords}
Fluid antenna system, channel estimation, spatial equalization, compressed sensing, group sparsity.
\end{IEEEkeywords}

\IEEEpeerreviewmaketitle

\section{Introduction\label{section 1}}
The increasing demands of 6G wireless networks, such as terabit-per-second data rates, centimeter-level positioning accuracy, and ultra-reliable low-latency communications, are fundamentally reshaping antenna system design. Conventional massive MIMO architectures, while successful in past generations' deployments, face inherent limitations imposed by their fixed half-wavelength spacing configurations. 

The Fluid Antenna System (FAS)~\cite{wong2020fluid,wong2021fluid,wong2022bruce} is a promising approach that exploits spatial diversity within confined regions to enhance physical-layer degrees of freedom (DoF), enabling diverse performance gains~\cite{wong2020performance}.

As explained in~\cite{zhu2024historical}, FAS generally refers to any software-controllable fluidic, dielectric, or conductive structures, including but not limited to liquid-based antennas~\cite{huang2021liquid}, pixel-based antennas~\cite{zhang2024novel}, and metasurface-design antennas~\cite{hoang2021computational,wang2025electromagnetically}, that can dynamically reconfigure their shape~\cite{zhao20233}, position~\cite{mitha2021principles}, orientation~\cite{shao20256dma}, and other radiation characteristics~\cite{zheng2025tri}. 
Specifically, FAS enables active reshaping of spatial sampling patterns, effectively converting physical array reconfiguration into enhanced spatial DoF. This paradigm shift allows real-time adaptation to channel state information (CSI) for achieving high spatial diversity gain,  high  spectral efficiency, and robust massive connectivity~\cite{wong2020performance,wong2021fluid,lai2023performance,zhu2023modeling,hu2024intelligent}. 

Existing FAS studies critically depended on comprehensive CSI acquisition at every feasible antenna position ~\cite{xu2024capacity,new2023information,new2023fluid}. Since FAS antennas move continuously within a confined aperture, realizing full performance gains necessitates CSI knowledge at arbitrary locations or numerous preset sampling points spaced below half-wavelength intervals. Recent advances in CSI acquisition leveraged sophisticated signal processing techniques, including array processing~\cite{wang2019overview}, data-driven approaches~\cite{neumann2018learning}, Bayesian learning frameworks~\cite{cheng2022rethinking}, compressive sensing~\cite{choi2017compressed}, etc. Such approaches leveraged intrinsic CSI properties: statistical behavior, sparsity structures, and cross-domain correlations across spatial, spectral, and temporal dimensions. These techniques mostly focused on scenarios with fixed radiating elements' locations, generally spaced in half-wavelength intervals. Conversely, FAS-assisted wideband systems face amplified full-CSI acquisition challenges due to the necessity for high-dimensional frequency-space channel reconstruction from sparse spatial-spectral observations.

Recent research has dedicated significant attention to channel estimation and reconstruction (CER) techniques for narrowband FAS, aiming for full CSI acquisition.~\cite{new2024channel} addressed the CER using Nyquist sampling and maximum likelihood estimation methods, and~\cite{wang2023estimation} used the least square (LS) method to reconstruct full CSI. In~\cite{skouroumounis2022fluid}, a skip-enabled linear minimum mean square error (LMMSE) channel estimation scheme is proposed, leveraging spatial correlation to select a subset of ports for reduced pilot overhead. In~\cite{zhang2024successive}, a model-free solution based on Bayesian regression was proposed, which also exploits the strong spatial correlation of the dense port of FAS. By exploiting the channel structure and the sparsity, the compressive sensing (CS) methods were also proposed for further reducing the pilot overhead and the spatial sampling, where both~\cite{ma2023compressed,xiao2024channel,cao2024channel} utilized the classical orthogonal matching pursuit (OMP) algorithm to deal with the sparse CER.~\cite{xu2023channel} separately estimated the angles of arrival (AoAs) and angles of departure using a combined discrete Fourier transform and rotation compensation method.~\cite{zhang2024channel} obtained the tensor's factor matrices via canonical polyadic decomposition to estimate angle and gain parameters, enabling channel reconstruction. Data-driven methods were also proposed to balance the accuracy and the training overhead~\cite{ji2024correlation,tang2025accurate}.
An important observation is that,
existing studies focused exclusively on narrowband FAS and performed CER solely in the spatial domain. 

Channel estimation for wideband FAS introduces a new dimension in the frequency domain. Particularly in environments with severe multipath propagation, wireless channels tend to exhibit frequency-space doubly-selective characteristics~\cite{bajwa2008learning,srivastava2021sparse}. Previous work  has also exploited this kind of sparse multiple domain joint channel estimation, including the delay-Doppler domain~\cite{wei2022off} and even delay-Doppler-angle domain~\cite{shen2019channel}. The core methodology for addressing these challenges lies in exploiting sparsity across joint domains to recover sparse parameters under the framework of CS.
However, for wideband FAS channel estimation, a fundamental distinction from conventional sparse recovery lies in the highly coherent dictionary—a CS concept—originating from the combined effects of limited aperture and oversampled ports. In other words, technically, we have to express the channel in wavenumber domain with a non-orthogonal basis or coherent dictionary~\cite{candes2011compressed,eiwen2010group,taubock2010compressive}. The coherent causes strong energy leakage and ambiguity in sparse domain, which exhibit a \textit{group sparse structure} for the wideband FAS channel. Although group sparse compressed sensing methodologies have been studied~\cite{eldar2009robust,eldar2010block,swirszcz2009grouped}  with proposed variants like group orthogonal matching pursuit (GOMP), their theoretical recovery objectives target sparse coefficient domain representations (delay-wavenumber domain), not signal domain expressions (frequency-space domain)~\cite{candes2011compressed}. 
Unlike spatial equalization in narrowband FAS systems, where deep fades can be mitigated by antenna relocation, in wideband scenarios, deep fades shift across the operating bandwidth rather than disappear\footnote{\label{foot:improvement by FAS} Note that the inherent frequency-space coupling phenomenon in wideband communication systems imposes more challenges compared to conventional  narrowband systems.
Traditional half-wavelength antenna arrays cannot effectively mitigate multiple deep in-band fading effects, particularly for subcarriers with sub-noise floor gain, especially in environments exhibiting significant delay spread and rich multipath components.
The electromagnetic equivalence between spatial displacement and temporal delay of path propagation reveals that sub-wavelength antenna repositioning can significantly reconfigure multipath interference patterns~\cite{dong2024wireless}.
The FAS can overcome this challenges in wideband system through real-time position adaptation within compact apertures, thereby providing extra spatial diversity comparing to fixed wideband array system.}. 
Consequently, conventional position optimization schemes inherently fail to eliminate deep fades throughout the entire bandwidth~\cite{xiao2024channel,chai2022port,xu2024capacity,xiao2024multiuser}. 
This fundamental limitation motivates us to propose a wideband FAS spatial equalization method to complete our joint space-frequency channel estimation and spatial equalization framework.

\begin{figure*}[ht]
    \centering
        \includegraphics[width=0.85\linewidth]{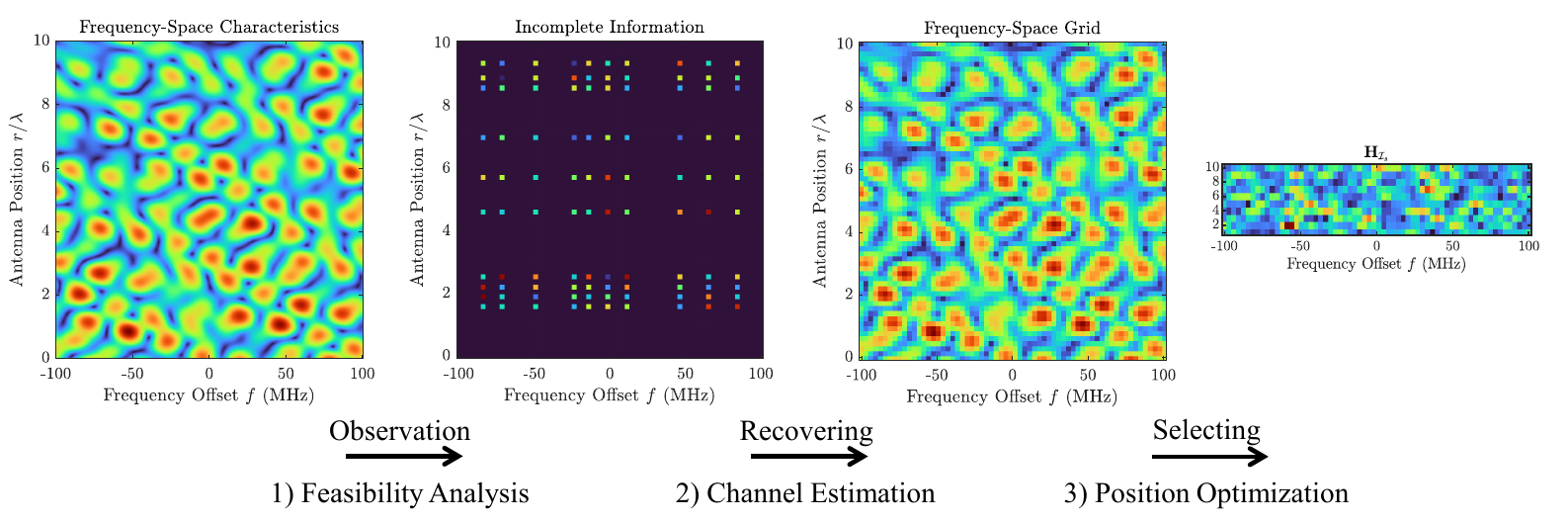}
        \caption{The tripartite methodology of this paper. (1) recovery feasibility analysis, (2) frequency-space channel estimation, and (3) positions optimization for spatial equalization.}
    \label{Procedure}
\end{figure*}
This paper, as shown in Fig.~\ref{Procedure}, aims to solve the above question through a tripartite methodology:
\begin{enumerate}
    \item \textit{Recovery feasibility analysis}: Establishing the theoretical support for the recovery of frequency-space characteristics (FSC) from limited observations, ensuring that the problem is well-posed and solvable under specific conditions. We then find that the physical aperture constraints in FAS induce leakage effects that degrade channel sparsity to a group-sparse structure, addressed by a novel convex optimization framework (CSOCP) minimizing $\ell_1/\ell_2$-mixed norms in delay-wavenumber domain. Theoretical  tight recovery bounds under dictionary-adapted group restricted isometry property (D-GRIP) conditions are also proposed, rigorously guaranteeing precise frequency-space channel reconstruction.
    \item \textit{Channel estimation}: Developing a computationally efficient algorithm to accurately reconstruct FSC from compressed observations, leveraging advanced optimization techniques and prior knowledge of signal structure. Based on the leakage-aware group-sparse theory, the descending correlation group orthogonal matching pursuit (DC-GOMP) is proposed whose dynamic descending-order grouping mechanism replaces static leakage-based group selection. This framework fundamentally overcomes subcoherence limitations, achieving order-of-magnitude efficiency gains in solving CSOCP problems through descending-correlative group orthogonalization.
    \item \textit{Positions optimization}: Designing an fast greedy algorithm to eliminate all the deep-fading within wideband, maximizing the minimal subcarrier gain.
    We reformulate the max-min antenna position optimization of SE as a mixed-integer linear programming (MILP) problem with guaranteed global optimality by the branch-and-bound method. To enable the real-time implementation, we further propose greedy row selection with isolated preselection (GRSIP) algorithm, a low-complexity algorithm that maintains around 80\% of optimal performance on average while reducing computation time by two orders of magnitude.
\end{enumerate}

Simulation results indicate that, (i) the proposed DC-GOMP greatly outperforms the conventional OMP and GOMP in lower-power details of recovery and convergence rate; (ii) The SE framework demonstrates scalable performance improvements in both deep fading suppression and average SNR enhancement, with efficacy positively correlated to aperture size configurations.

The rest of this paper is organized as follows. Section~II introduces the system model and formulation of channel estimation. In Section~III, the key theoretical framework based on the fact of leakage has been established.   Section~IV  gives  the   algorithms to perform the channel estimation and spatial estimation. In Section~V, numerical results are provided, and  the conclusion is drawn in Section~VI.

\textit{Notation}: The notation adopted in the paper is as follows. We use bold to denote matrices and vectors. Specifically, $\mathbf{I}_N$ denotes the $N\times N$ identity matrix. $\{\cdot\}^T$ and $\{\cdot\}^H$ denote the transpose, and the Hermitian operator. $\{\mathbf{X}\}^{\dagger}$ is the Moore-Penrose pseudoinverse of $\{\mathbf{X}\}$.  $\mathbf{\hat{X}}$ and $\mathbf{\hat{v}}$ denote the estimators of matrix $\mathbf{X}$ and vector $\mathbf{v}$. $\operatorname{vec}(\cdot)$ denotes the vectorization of matrix. $\textbf{blkdiag}\{\mathbf{A}_1,\dots,\mathbf{A}_N\}$ denotes a block diagonal matrix with $\mathbf{A}_1,\dots,\mathbf{A}_N$ at the main diagonal. $\text{Rank}\{\mathbf{X}\}$ and $\text{tr}\{\mathbf{X}\}$ denote the rank and trace of matrix $\mathbf{X}$. $\Re\{\cdot\}$ and $\Im\{\cdot\}$ respectively denote the real and image part of a complex number.

\section{System Model\label{section 2}}
In this section, we first provide the definition of the space-frequency characteristics (SFC) of FAS and its wideband multipath propagation modeling in the space-frequency domain. Then we transform the SFC into the sparse delay-wavenumber domain and indicate the leaky phenomenon caused by limited aperture size and bandwidth. Then we propose the observation model with limited numbers of antennas and pilots. Finally, we propose the two-step framework for this model as depicted in Fig.~\ref{Procedure}, with the objective to maximize spatial diversity gain. 
\subsection{Doubly-Selective Space-Frequency Characteristics\label{section 2A}}
 The SFC plays a critical role in designing the optimal antenna positions within a FAS. The wave propagation between two points comprises multiple propagation paths, each with distinct delays, angles, and signal strengths. Due to frequency dispersion, the propagation experiences varying delay combinations across different frequencies, leading to spectral selectivity. As an antenna moves within its available aperture—measured in units of the reference wavelength—the spectral selectivity undergoes significant variations. The SFC encapsulates all possible propagation modes between the regions  $\mathcal{S}_{tx}$ and $\mathcal{S}_{rx}$ 
  over the bandwidth 
$\mathcal{B}$, given specific boundary conditions and propagation environments. Fundamentally, it represents a simplified form of the Green's function, characterizing the system's response to different spatial and frequency-dependent variations.
\begin{definition}[SFC]
\label{def:FSCF}
The SFC represents the mapping from a compact transmit region $\mathcal{S}_{\text{tx}} \subset \mathbb{R}^3$ to a compact receive region $\mathcal{S}_{\text{rx}} \subset \mathbb{R}^3$ in a wireless communication system with spectral support $\mathcal{B} \subset \mathbb{R}$. It is defined as the kernel of a compact operator $\mathcal{G}$:
\begin{equation}
(\mathcal{G}\mathbf{x})(\mathbf{p}_{\text{rx}}, \omega) = \int_{\mathcal{S}_{\text{tx}}} g(\mathbf{p}_{\text{rx}}, \mathbf{p}_{\text{tx}}, \omega) x(\mathbf{p}_{\text{tx}}, \omega) \, d\mathbf{p}_{\text{tx}},
\label{eqFSCF}
\end{equation}
where $\omega \in \mathcal{B}$ denotes angular frequency of the propagating wave, $x(\mathbf{p}_{\text{tx}}, \omega): \mathcal{S}_{\text{tx}} \times \mathcal{B} \to \mathbb{C}$ is the source density function.\footnote{\label{foot: the source density function}For example, in electromagnetic systems, $x(\mathbf{p}_{\text{tx}}, \omega)$ represents current density (A/m$^2$) or charge distribution (C/m$^3$)} $\mathbf{p}_{\text{rx}} \in \mathcal{S}_{\text{rx}}$ denotes receiver coordinates. $g(\mathbf{p}_{\text{rx}}, \mathbf{p}_{\text{tx}}, \omega)$ encodes wave propagation characteristics.
\end{definition}

We consider the case that the transmitting region $\mathcal{S}_{tx}$ has only one spatial point, the receiving region $\mathcal{S}_{rx}$ consists of a line segment of a size $\vert\mathcal{S}_{rx}\vert=W$, and limited bandwidth $\mathcal{B}=B$. This allows us to simplify the notation $x(\mathbf{p}_{tx}, \omega)$ as $x(\omega)$; thus the general  SFC function in (\ref{eqFSCF}) is reduced to
\begin{equation}
    (\mathcal{G}x)(r,\omega)=g(r,\omega)x(\omega),
    \label{FSCF}
\end{equation}
where  $\omega\in[-B/2,B/2]$ and $r\in[0,W]$ represent  the frequency and space components, respectively. 
Frequency dispersion is a phenomenon in which the propagation characteristics of a channel or medium vary with frequency, resulting in different frequency components traveling at distinct phase velocities or experiencing different attenuation levels. 
For a specific path from the fixed transmitting antenna (Tx) to the receiving antenna (Rx) with coordinate $\mathbf{p}$, its frequency response can be expressed as $\alpha_l(\omega)e^{j\left( \mathbf{k}_l^T\mathbf{p}+\omega\tau_l\right)}$. The attenuation that varies with frequency is typically taken into account in ultra-wideband wireless channels~\cite{qiu1999multipath}\cite{qiu2002study}. 
In the case of wideband, we disregard the frequency-dependent attenuation for each path and only consider the frequency-dependent array manifold. 
The 
channel response 
from the fixed source point to a movable sink point at position $r$ and frequency $\omega$  can be expressed as
\begin{equation}
    g(r,\omega)=\sum_l^L\alpha_le^{j\frac{\omega+\omega_c}{c}\left(r\text{cos}\theta_l+c\tau_l\right)},
    \label{wireless P2P channel}
\end{equation}
which presents the sum of $L$ resolvable paths with different angle-of-arrival (AoA) $\theta_l\in(0,\pi ]$ and group delays $\tau_l\in[0, \tau_{max}]$. The complex attenuation $\alpha_l$ is the result of the small scale fading that satisfy the normalized Rayleigh distribution, i.e., $\alpha_l \stackrel{\text { i.i.d }}{\sim} \mathcal{CN}(0,1),\forall l$. 
The $\omega_c$ denotes the angular frequency of the central carrier.

Denote the number of considered discrete positions   in $[0,W]$ by  $M$,   and the number of considered discrete frequencies   in $[-B/2,B/2]$ by $K$. 
We obtain the discrete expression of $g(r,\omega)$ by the resolutions of $\Delta r\ll \lambda$ and $\Delta \omega \leq 2\pi/\tau_{\text{max}}$ in the space and frequency domains\footnote{\label{foot:taumax}$\tau_{\text{max}}$ is the maximal delay spread of the wireless propagation.}
which is denoted as space-frequency grid (SFG) $\mathbf{G}\in\mathbb{C}^{M\times K}$,
\begin{equation}
    \mathbf{G}=[\mathbf{g}_1,\mathbf{g}_2,\dots,\mathbf{g}_K]=[\bar{\mathbf{g}}_1,\bar{\mathbf{g}}_2,\dots,\bar{\mathbf{g}}_M]^T,
    \label{spatial selectivity profile}
\end{equation}
 where the vector $\mathbf{g}_k\in\mathbb{C}^{M\times 1}$ denotes the space-selective response  at $k$-th frequency point $(\frac{k}{K}-\frac{1}{2})B$, and the $m$-th element of $\mathbf{g}_k$ is  $\mathbf{g}_k(m)=g\left(\frac{m}{M}W,(\frac{k}{K}-\frac{1}{2})B\right)$; the vector $\bar{\mathbf{g}}_m\in\mathbb{C}^{K\times 1}$ denotes the frequency-selective response  at $m$-th location point in space domain.

\subsection{Sparse Representation in Delay-Wavenumber Domain\label{section 2B}}
\label{sec:sparse representation}
For a FAS with finite aperture size and bandwidth, the SFG can be represented sparsely in the  delay-wavenumber domain as Dirac-like functions with respect to wavenumber and delay of each path.
Although different frequency components from each path have different phase shifts, they share the same AoA (i.e., they share the same angular support), which can be depicted as the expression in the wavenumber-frequency domain as
\begin{equation*}
    \tilde{g}(\mathbf{k},\omega)=\sum_{l=1}^L\alpha_l e^{j(\omega+\omega_c)\tau_l}W\operatorname{sinc}(\frac{W(\omega+\omega_c)(\mathbf{k}_l-\mathbf{k})}{2c}),
\end{equation*}
where $\mathbf{k}_l=\cos\theta_l$ and $\theta_l$ denote the wavenumber and the AoA of $l$-th path, respectively. 
By taking the Inverse Fourier Transform  of $  \tilde{g}(\mathbf{k},\omega)$ on  $\omega$, we have the expression in the delay-wavenumber domain as
\begin{equation}
     \tilde{\tilde{g}}(\mathbf{k},\tau)\approx WB\sum_{l=1}^L\alpha_le^{j\omega_c(\tau_l-\tau)}\mathbf{\Lambda}_l(\mathbf{k},\tau),
\end{equation}
where the leakage pattern can be described as
\begin{equation}
 \mathbf{\Lambda}_l(\mathbf{k},\tau)=\operatorname{sinc}(\frac{W(2\omega_c-B)(\mathbf{k}_l-\mathbf{k})}{4c})\operatorname{sinc}(\frac{B(\tau_l-\tau)}{2}).
    \label{leakage pattern}
\end{equation}
The leakage effect implies a degraded sparsity in the wavenumber domain~\cite{taubock2010compressive}. To study this relationship, we consider the energy decay of the leakage pattern; given a specific path, the amplitude decay has the envelope $\vert \mathbf{\Lambda}_l(\mathbf{k},\tau) \vert\leq \mathbf{\Lambda}_l^{\textbf{env}}(\mathbf{k},\tau) :=8c(WB(2\omega_c-B)\vert\mathbf{k}_l-\mathbf{k}\vert\vert\tau_l-\tau\vert)^{-1}$, where $\tilde{\tilde{g}}(\mathbf{k},\tau)$ can be considered as approximately sparse (or compressible) in the wavenumber domain~\cite{candes2006stable}.
As a result, a leakage coefficient $\gamma$ can be obtained as  follows.
\begin{definition}
        Given the aperture size $W$ and bandwidth $B$ of the FAS and the power detection threshold $T$. For an arbitrary frequency bin, the leakage coefficient in wavenumber domain can be defined as
    \begin{equation}
        \gamma=\frac{\vert \{(\mathbf{k},\tau)\vert\mathbf{\Lambda}^{\textbf{env}}_l(\mathbf{k},\tau)^2\ge T\} \vert}{\Delta\mathbf{k}\Delta\tau},
        \label{leakage}
    \end{equation}
    where  $\Delta\mathbf{k}$ and $\Delta\tau$ denote  the resolution of the sufficient grid of $\tilde{\tilde{g}}(\mathbf{k},\tau)$. Normally, $T$ is larger than the noise power $\sigma_{\omega}^2$.
    \label{degradation}
\end{definition}

From the propagation perspective, the sparsity of the wavenumber domain depends on the number of paths $L$. The support set of these paths is identical and corresponds to the set of indices in the wavenumber domain associated with each path, i.e., $\operatorname{supp}\{\tilde{\mathbf{g}}_k\}=\{\mathbf{k}_l/\mathbf{k}\}_L,\forall k$. We can express the spatially-selective profile at $k$-th frequency point as 
\begin{equation}
    \mathbf{g}_{k} = \mathbf{A}(\omega_k)\mathbf{\tilde{g}}_{k},
    \label{transform}
\end{equation}
where the array manifold matrix $\mathbf{A}(\omega_k)\in\mathbb{C}^{M\times M}$ maps the wavenumber domain into the space domain, 
\begin{equation}
    \begin{aligned}
        \mathbf{A}(\omega_k)=\left[ \mathbf{a}(\mathbf{k}_1;\omega_k),\mathbf{a}(\mathbf{k}_2;\omega_k),\dots,\mathbf{a}(\mathbf{k}_M;\omega_k) \right].
    \end{aligned}
    \label{array manifold matrix}
\end{equation}
Note that $\{\mathbf{k}_q\}_M$ is the uniform sampling points in wavenumber domain and $\mathbf{a}(\mathbf{k}_l;\omega_k)\in\mathbb{C}^{M\times 1}$ is the frequency-dependent array manifold,
\begin{equation}
    \mathbf{a}(\mathbf{k};\omega_k)=\left[1,e^{j\frac{W(\omega_k+\omega_c)\mathbf{k}}{cM}},\dots,e^{j\frac{(M-1)W(\omega_k+\omega_c)\mathbf{k}}{cM}}\right]^T.
\end{equation}

\subsection{Compressed Observation Model\label{section 2C}}
Considering the propagation model of SFG and the intrinsic nature of the transformation, we give the noisy compressed observation model in this subsection.
Let us consider the wideband SIMO system where the movable array has the initial set of antenna positions $\boldsymbol{\mathcal{I}_r}\subset\{1, 2, \dots, M\},\vert\boldsymbol{\mathcal{I}_r}\vert=N_r$, and inserted pilots have positions in the set $\boldsymbol{\mathcal{I}_p}\subset\{1, 2, \dots, K\},\vert\boldsymbol{\mathcal{I}_p}\vert=N_p$. 
The pilot located at the $i\in\boldsymbol{\mathcal{I}_p}$-th frequency bin is denoted as $s_i\in\mathbb{C}$.
Here we define the location-selecting matrix $ \mathbf{S_{\boldsymbol{\mathcal{I}_r}}}\in\mathbb{C}^{N_r\times M}$ and the pilots-selecting matrix $\mathbf{S_{\boldsymbol{\mathcal{I}_p}}}\in\mathbb{C}^{N_p\times K}$ as 
\begin{equation}
\begin{aligned}
    \mathbf{S_{\boldsymbol{\mathcal{I}_r}}}&=[\mathbf{e}_{i_{r,1}},\dots,\mathbf{e}_{i_{r,N_r}}]^T, i_{r,n}\in\boldsymbol{\mathcal{I}_r}; \\
    \mathbf{S_{\boldsymbol{\mathcal{I}_p}}}&=[s_{i_{p,1}}\bar{\mathbf{e}}_{i_{p,1}},\dots,s_{i_{p,N_p}}\bar{\mathbf{e}}_{i_{p,N_p}}]^T,i_{p,n}\in\boldsymbol{\mathcal{I}_p}.
    \label{sample matrix}
\end{aligned}
\end{equation}
Note that $\mathbf{e}_{i_{r,n}},n=1,\dots,N_r$ (each with dimension $M\times 1$) and $\bar{\mathbf{e}}_{i_{c,n}},n=1,\dots,N_c$ (each with dimension $K\times 1$) represent the $n$-th standard unit vectors, where the $n$-th element is $1$ and all other elements are $0$. We can model the available observations of SFG at the receiver as 
\begin{equation}
    \mathbf{Y}_{\text{m}}= \mathbf{S_{\boldsymbol{\mathcal{I}_r}}}\mathbf{G}\mathbf{S}^T_{\boldsymbol{\mathcal{I}_p}}+\mathbf{Z},
    \label{selecting}
\end{equation}
where $\mathbf{Z}\in\mathbb{C}^{N_r\times N_c}$ denotes the i.i.d. noise matrix with covariance matrix $\sigma^2_w\mathbf{I}$, and $\mathbf{Y}_{\text{m}}\in\mathbb{C}^{N_r\times N_c}$ denotes the incomplete observations of the SFG.
 Due to the frequency dispersion, the array manifold matrix varies as the frequency changes, and we need to apply the vectorization expression for all $K$ frequency points within the bandwidth. We impose the operator $\mathrm{vec(\cdot)}$ on both sides of~\eqref{selecting},
\begin{equation}
    \operatorname{vec}(\mathbf{Y}_{\text{m}})=\left(  \mathbf{S_{\boldsymbol{\mathcal{I}_p}}}\otimes\mathbf{S_{\boldsymbol{\mathcal{I}_r}}}\right)\operatorname{vec}(\mathbf{G})+\mathbf{z},
\end{equation}
where $\otimes$ denotes the Kronecker product, $\mathbf{z}\in\mathbb{C}^{N_rN_p\times 1}$ is the i.i.d. Gaussian noise vector, and the identical equation $\operatorname{vec}(\mathbf{C})=\operatorname{vec}(\mathbf{A} \mathbf{X} \mathbf{B})=\left(\mathbf{B}^{\top} \otimes \mathbf{A}\right) \operatorname{vec}(\mathbf{X})$ is used. The frequency-space domain $\operatorname{vec}(\mathbf{G})\in \mathbb{C}^{MK\times 1}$ can be expressed as the linear transform of the sparse matrix $\mathbf{\tilde{G}}\in\mathbb{C}^{M\times K}$ in frequency-angular domain.
\begin{equation}
\begin{aligned}
    \operatorname{vec}(\mathbf{G})
    =\textbf{blkdiag}\{\mathbf{A}\left(\omega_1\right),\dots,\mathbf{A}\left(\omega_K\right)\}\operatorname{vec}(\mathbf{\tilde{G}}).
\end{aligned}
\end{equation}
Note that $\mathbf{\tilde{G}}=\mathbf{\tilde{\tilde{G}}}\mathbf{F}$ where $\mathbf{F}$ is the DFT matrix satisfying $\mathbf{F}^H\mathbf{F}=\mathbf{I}_K$. By $\operatorname{vec}(\mathbf{A} \mathbf{B})=\left(\mathbf{B}^{\mathrm{T}} \otimes \mathbf{I}_M\right) \operatorname{vec}(\mathbf{A})$, we have 
\begin{equation}
    \operatorname{vec}(\mathbf{\tilde{G}})=\left(\mathbf{F}^{\mathrm{T}} \otimes \mathbf{I}_M\right) \operatorname{vec}(\mathbf{\tilde{\tilde{G}}}).
\end{equation}
Based on the compressed sensing process~\cite{candes2011compressed}, we can model the relationship of observation, SFG and frequency-angular grid as (\ref{sparse model}) , where $\mathbf{y_0}=\operatorname{vec}(\mathbf{Y}_{\text{m}})\in\mathbb{C}^{N_rN_c\times 1}$ denotes the observations, $\mathbf{x_0}=\operatorname{vec}(\mathbf{\tilde{\tilde{G}}})$ denotes the sparse vector. $\mathbf{S}=\mathbf{S_{\boldsymbol{\mathcal{I}_p}}}\otimes\mathbf{S_{\boldsymbol{\mathcal{I}_r}}}\in\mathbb{C}^{N_rN_c\times MK}$ and $\mathbf{\Omega}=\textbf{blkdiag}\{\mathbf{A}\left(\omega_1\right),\dots,\mathbf{A}\left(\omega_K\right)\}\in\mathbb{C}^{MK\times MK}$. 
$\mathbf{\Psi}=\left(\mathbf{F}^{\mathrm{T}} \otimes \mathbf{I}_M\right)\in\mathbb{C}^{MK\times MK}$ represents the DFT from the sparse delay-wavenumber domains into group-sparse frequency-angular domain satisfying $\mathbf{\Psi}^H\mathbf{\Psi}=\mathbf{I}_{MK}$.

\begin{figure*}[tp]
    \centering
    \begin{equation}
    \underbrace{\operatorname{vec}\left(\mathbf{Y}_{\text {m }}\right)}_{\mathbf{y}_0}=\underbrace{\underbrace{\left(\mathbf{S_{\boldsymbol{\mathcal{I}_p}}} \otimes \mathbf{S_{\boldsymbol{\mathcal{I}_r}}}\right)}_{\mathbf{S}} \underbrace{\mathbf{b l k d i a g}\left\{\mathbf{A}\left(\omega_1\right), \ldots, \mathbf{A}\left(\omega_K\right)\right\}\left(\mathbf{F}^T \otimes \mathbf{I}_M\right)}_{\mathbf{D}}}_{\mathbf{M}} \underbrace{\operatorname{vec}(\tilde{\tilde{\mathbf{G}}})}_{\mathbf{x}_0}+\mathbf{z}.
    \label{sparse model}
    \end{equation}
\end{figure*}

\subsection{Two-Step Framework}
\label{section 2D}
We develop a two-step framework for the FAS-SIMO: first resolving space-frequency channel reconstruction through compressed sensing, then executing antenna position optimization to maximize spatial diversity gain.

{\bf  Channel Estimation and Reconstruction.}
The recovery of the SFG at the receiver  is equivalent to the space-frequency channel estimation of the SIMO system with a FAS-assisted receiver.
Due to the aperture limitation and the high demand of the angular/spatial resolution, as mentioned in Section~\ref{sec:sparse representation}, the  severe leakage effect and coherence between adjacent grids in SG are inevitable.  Lemma~\ref{degradation} shows that $\|\mathbf{\Psi}^H\mathbf{\Omega}^H\operatorname{vec}(\mathbf{G})\|^2_2$ decays rapidly at the order of $-2$, and $\mathbf{\Psi}^H\mathbf{\Omega}^H\mathbf{\Omega}\mathbf{\Psi}$ is a well-behaved block-diagonal sparse matrix.

In this study, we do not aim to recover the exact wireless propagation in delay-wavenumber domain but the SFC of FAS which can be sufficiently regarded as SFG (i.e., $\mathbf{G}$). We can obtain the estimated SFG $\mathbf{\hat{G}}$ from the available observations in (\ref{selecting}) by the method of $\ell_1$-analysis,
\begin{subequations}
\label{SFG recovery}
\begin{align}
        \underset{\mathbf{G} \in \mathbb{C}^{M\times K}}{\arg \min}
        \ &\|\mathbf{\Psi}^H\mathbf{\Omega}^H\operatorname{vec}(\mathbf{G})\|_1 \\
        \text{s.t.}\ &\| \mathbf{S}\operatorname{vec}(\mathbf{G}) - \mathbf{y}_{\text{m}} \|_2\leq\varepsilon,
\end{align}
\end{subequations}
where $\varepsilon$ denotes an upper bound on the noise level $\sigma_{\omega}$. The feasibility and stability of (\ref{SFG recovery}) have been analyzed, and more comprehensive conditions have been established to ensure the effective performance of the $\ell_1$-analysis~\cite{candes2011compressed}.

{\bf Spatial Equalization} 
In order to mitigate deep fading, we obtain the optimal positions $\boldsymbol{\mathcal{I}_s}$ to maximize the gain of the subcarrier with the minimal power by the recovered SFG.
The wideband receiving signal can be expressed as
\begin{equation}
    \mathbf{H}_{\boldsymbol{\mathcal{I}_s}}=[\mathbf{\bar{g}}_{i_1}, \mathbf{\bar{g}}_{i_2},\dots,\mathbf{\bar{g}}_{i_{N_r}}]^T=[\mathbf{h}_1,\mathbf{h}_2,\dots,\mathbf{h}_K],
\end{equation}
where $\mathbf{H}_{\boldsymbol{\mathcal{I}_s}}\in\mathbb{C}^{N_r\times K}$ and $i_n\in\boldsymbol{\mathcal{I}_s}$.  The $k$-th subcarrier of the receiving signal can be expressed as
\begin{equation}
     y_k=\mathbf{w}_k^H(\mathbf{h}_k x_k+\mathbf{z}),
\end{equation}
where $\mathbf{w}_k\in\mathbb{C}^{N_r\times K}$ is the combining weights of $k$-th subcarrier and $\mathbf{h}_k$ is the $k$-th column of $\mathbf{H}_{\boldsymbol{\mathcal{I}_s}}$. The $n$-th row of $\mathbf{H}_{\boldsymbol{\mathcal{I}_s}}$ (i.e., $\mathbf{\bar{g}}_{i_n}$) is equal to $i_n$-th row of $\mathbf{G}$. The $\boldsymbol{\mathcal{I}_s}^{\text{opt}}$ can be obtained by
\begin{subequations}
\label{position optimation1}
\begin{align}  
\underset{\boldsymbol{\mathcal{I}_s}}{\arg \max}\ \underset{k}{\min} &\  \rho_k\\
        \text{s.t.}&\  \rho_k=\mathbb{E}\{|\mathbf{w}_k^H\mathbf{h}_k x_k|^2\}/\mathbb{E}\{ |\mathbf{w}_k^H\mathbf{z}|^2\},\\
        &\ h_{k,i_n}= G_{i_n,k},\ i_n\in \boldsymbol{\mathcal{I}_s},\\
        &\ \vert\boldsymbol{\mathcal{I}_s}\vert=N_r,  \boldsymbol{\mathcal{I}_s}\subset \{1,\dots,M\},
\end{align}
\end{subequations}
where $G_{i_n,k}$ is the entries of $\mathbf{G}$ at $i_n$-th row and $k$-th column ($k \in \{1,\ldots,K\}$), and $\rho_k$ is the average SNR of the $k$-th subcarrier. To counteract the selectivity and guarantee the robustness of the transmission, we apply the MRC for $k$-th subcarrier along with each antenna, i.e., $\mathbf{w}_k=\mathbf{h}_k$. 

\section{Channel Estimation based on Group Sparsity}
\label{section III}
This section first presents a theoretical analysis of error bounds for group sparsity recovery of SFG. Under the noisy environments with leakage effects, sparse signal recovery faces two critical challenges: (i) the degradation of sparsity in the delay-angle domain due to leakage, where non-zero elements cluster into group structures, and (ii) the coherence of the measurement matrix $\mathbf{M}$, which destabilizes recovery. To analyze the feasibility, we introduce the D-GRIP and establish a rigorous upper bound on the SFG recovery error. This bound depicts the relationship between recovery error, group sparsity level $k$, number of observations $N_rN_p$, and dictionary redundancy.  
Furthermore, we define group coherence $\mu_{\mathcal{J}}$ and sub-coherence $\nu$, characterize the group structure under leakage effects, and derive a theoretical upper bound on group size $\gamma$,   providing theoretical guidelines for designing channel estimation algorithms. 

Subsequently, we design an efficient and robust algorithm, called \textit{descending correlation group orthogonal matching pursuit} (DC-GOMP), based on the group sparsity and the fact that the subcoherence inside one leakage group is high. 

\subsection{Error Bound of Group Sparsity Recovery\label{section 3A}}
\label{sec:error bound}
We  use the model in (\ref{sparse model}) to explain the concept of group sparsity. The noisy observation model is
\begin{equation}
\mathbf{y_0}=\mathbf{M}\mathbf{x_0}+\mathbf{z}=\mathbf{Sg_0}+\mathbf{z}=\mathbf{S}\mathbf{D}\mathbf{x_0}+\mathbf{z},
    \label{observation model}
\end{equation}
where $\mathbf{M}=\mathbf{SD}=\mathbf{S\Omega}\mathbf{\Psi}\in\mathbb{C}^{N_rN_c\times MK}$ with $N_rN_c \ll MK$, and $\mathbf{x_0}$ denotes vectorized sparse expression in the delay-wavenumber domain. From the perspective of propagation, $\mathbf{x_0}$ should be $L$-sparse corresponding to the number of paths. Nonetheless, as outlined in Section~\ref{sec:sparse representation}, the leakage effect diminishes its sparsity characterized by a factor $\gamma$, resulting in $\gamma L$ non-zero elements emerging in groups. Let $\mathcal{J} \triangleq\left\{\mathcal{I}_j\right\}_{j=1}^J$ be a partition of the set $\{1, \dots,MK\}$, i.e., $\bigcup_{j=1}^J \mathcal{I}_j=\{1, \dots,MK\}$ and $\sum_{j=1}^J\vert\mathcal{I}_j\vert=MK$. We assign the size of each group $\vert\mathcal{I}_j\vert=\gamma$ to match the sparsity degradation. 
For the $\mathbf{x_0}\in\mathbb{C}^{MK\times 1}$ and $\mathbf{M}\in\mathbb{C}^{N_rN_c\times MK}$, we define the $j$-th subvector of size $\gamma \times 1$ 
\begin{equation}
    \mathbf{x_0}[j]=[\tilde{\tilde{x}}_{i}: i\in\mathcal{I}_j  ]^T,
\end{equation}
and $j$-th column-group of size $N_rN_c\times\gamma$ 
\begin{equation}
    \mathbf{M}[j]=[\mathbf{m}_{i}: i\in\mathcal{I}_j],
\end{equation}
where $\mathbf{m}_{i}$ is the $i$-th column of $\mathbf{M}$. A vector $\mathbf{x_0}$ is called \textit{group $L$-sparse} as the condition $\sum_{j=1}^J I(\|\mathbf{x_0}[j]\|_2)\leq L$ holds, where $I(\cdot)$ is the indicator function.

By defining the $\ell_1/\ell_2$-norm $\|\cdot\|_{2,\mathcal{J}}$ where $\|\mathbf{x_0}\|_{2,\mathcal{J}}=\sum_{j=1}^J \|\mathbf{x_0}[j]\|_2$, we now rewrite the optimization problem in~\eqref{SFG recovery} as the CSOCP,
\begin{subequations}
\label{SFG group recovery}
\begin{align}
\hat{\mathbf{g}}=\underset{\mathbf{g} \in \mathbb{C}^{M K \times 1}}{\arg \min }\ &\left\|\mathbf{D}^H \mathbf{g}\right\|_{2, \mathcal{J}} \\
\text { s.t. }\ &\left\|\mathbf{S g}-\mathbf{y}_{\mathbf{0}}\right\|_2 \leq \varepsilon .
\end{align}
\end{subequations}
 We will then impose a natural group property on the sampling matrix $\mathbf{S}$, analogous to the group restricted isometry property defined in ~\cite{eldar2009robust,candes2011compressed}.
\begin{definition} (D-GRIP) 
Let $\Sigma_k$ be the space spanned by  given dictionaries $\mathbf{D}$ and   group $k$-sparse vector $\mathbf{x}$, i.e., $\mathbf{Dx}\in\Sigma_k$, $\forall \mathbf{x},\ \text{s.t.}\ \|\mathbf{x}\|_{2,0,\mathcal{J}}\leq k$. The sampling matrix $\mathbf{S}$ is said to obey the \textit{group restricted isometry property} adapted to $\mathbf{D}$ (abbreviated D-GRIP) with constant $\delta_k$ if
    \begin{equation}
        (1-\delta_k)\|\mathbf{v}\|_2^2\leq\|\mathbf{Sv}\|_2^2\leq(1+\delta_k)\|\mathbf{v}\|_2^2
    \end{equation}
    holds for all $\mathbf{v}\in\Sigma_k$.
\end{definition}
We point out that $\Sigma_k$ is just the image under $\mathbf{D}$ of all group $k$-sparse vectors. The D-GRIP degenerates into standard RIP when $\mathbf{D}=\mathbf{I}$~\cite{candes2006robust}.
\begin{theorem}(Error Bound)
    Let $\mathbf{y_0}=\mathbf{S}\mathbf{g_0}+\mathbf{z}=\mathbf{S}\mathbf{D}\mathbf{x_0}+\mathbf{z}$ be noisy observations of a group $k$-sparse vector and $\mathbf{D}$ is an  arbitrary tight frame. Let $\mathbf{\hat{g}}$ be a solution to (\ref{SFG group recovery}) and $(\mathbf{D}^H\mathbf{g_0})^{(k)}$ be the largest group $k$-sparse approximation of $\mathbf{D}^H\mathbf{g_0}$. If $\mathbf{S}$ satisfies the D-GRIP with $\delta_{k+P}<1$ then

    \begin{equation}
        \|\mathbf{\hat{g}}-\mathbf{g_0}\|_{2}\leq C_0\|\mathbf{D}^H\mathbf{g_0}-(\mathbf{D}^H\mathbf{g_0})^{(k)}\|_{2,\mathcal{J}}+C_1 \varepsilon,
    \end{equation}
    where 
\begin{equation}
    C_0=2\frac{a\sqrt{1-\delta_{k+P}}+\sqrt{1+\delta_P}}{b\sqrt{1-\delta_{k+P}}-\sqrt{1+\delta_P}},
\end{equation}
\begin{equation}
    \ C_1=\frac{2}{b\sqrt{\rho}\sqrt{1-\delta_{k+P}}-\sqrt{1+\delta_P}},
\end{equation}
and $a=\sqrt{1+\frac{1}{c}},\ b=\sqrt{\frac{1}{\rho}-1-c},\ \rho=k/P,$ with the positive number $c$.
\label{Error Bound}
\end{theorem}
\textit{Proof}: \textit{The proof of Theorem~\ref{Error Bound} is inspired by both the result of compressed sensing with redundant dictionaries~\cite{candes2011compressed} and the robust recovery of block sparse vector~\cite{eldar2009robust}. The new challenge here is that, rather than bounding the error $\|\mathbf{\hat{x}}-\mathbf{x_0}\|_2$ in group sparse domain with related to the mixed norm $\|\cdot\|_{2,\mathcal{J}}$, we try to bound the image error $\|\mathbf{\hat{g}}-\mathbf{g_0}\|_{2}$ under the projection $\mathbf{D}$. By bounding the group tail of $\|\mathbf{D}^H(\mathbf{\hat{g}}-\mathbf{g_0})\|_2$ (step 1) and the application of the D-GRIP (step 2), we finally obtain the expression of the error bound (step 4). See Appendix~\ref{sec:append proof of th1} for the details.}$\hfill\blacksquare$

Theorem~\ref{Error Bound} provides theoretical support for the SFG recovery problem by noting the group sparsity of SFG in the delay-wavenumber domain. Given the number of observations $N_r$ and $N_p$, the upper bound of recovery error is linked to the tail power $\|\mathbf{D}^H\mathbf{g_0}-(\mathbf{D}^H\mathbf{g_0})^{(k)}\|_{2,\mathcal{J}}$, noise power $\varepsilon$, or group sparsity $k$. This emphasizes choosing the correct group support for effective recovery, leading to a group-based algorithm proposed later. As the number of observations grows, the group restricted isometry constant adapted to $\mathbf{D}$ (i.e., $\delta_k$) decreases, reducing the error bound.
In the absence of noise and leakage, $\ell_1/\ell_2$-norm minimization or traditional greedy algorithms can accurately solve the problem in (\ref{SFG group recovery}). However, noise and restricted aperture size and bandwidth prevent perfect recovery.

The upper bound on the error established in Theorem~\ref{Error Bound} confirms that the problem~\eqref{SFG group recovery}  is feasible under the proper D-GRIP condition. However, the actual recovery error is highly dependent on the specific recovery algorithm employed. Unlike conventional sparse recovery or channel estimation, which aims for precise parameter estimation, our primary objective is to reconstruct the SFG, thereby simplifying the problem. In environments with a large number of propagation paths, the limited aperture size and bandwidth constrain the number of resolvable paths, which is often smaller than the actual number of physical paths. Paths that are close in the delay-wavenumber domain tend to merge due to leakage effects, resulting in indistinguishable components at the system level. 
Nevertheless, this negative effect can be mitigated since our recovery object is the expression in space-frequency domain rather than delay-wavenumber domain.
This distinction motivates the development of more targeted algorithms, which will be introduced in Section~\ref{sec:DC-GOMP}.

\subsection{Leakage and Group Sparsity of Channel}
We demonstrate—for the first time—a group-sparse structure in wideband aperture-confined FAS channels. Crucially, this group-sparsity pattern originates not from clustered scatterers (conventionally assumed in group-sparse channel models), but from the high coherence of adjacent steering vectors induced by spatially oversampled dense ports. Consequently, this introduces significant ambiguities in distinguishing the precise sparse signal in high-resolution delay-wavenumber grid.
Under the assumption of high space resolution, the array manifold matrix in~\eqref{array manifold matrix} is actually the oversampled discrete Fourier transform (DFT) matrix in which the sampled frequencies are taken over at small intervals much less than wavelength. This leads to an overcomplete frame whose columns are highly correlated, e.g., $\mathbf{a}(\mathbf{k}_{m};\omega_k)^T\mathbf{a}(\mathbf{k}_{m-1};\omega_k)\to 1$. 
The high coherence of the overcomplete frame and the leakage effect are fundamentally equivalent here, both arising from the inability of the aperture to achieve high angular resolution. This is analogous to the relationship between spectral bandwidth and sampling rate. When the sensing matrix (specifically, the oversampled array manifold matrix) exhibits high coherence, the resulting observations tend to be highly correlated. 
Consequently, this introduces significant ambiguities in distinguishing the precise sparse signal in high-resolution delay-wavenumber grid.
Fortunately, in this paper, the recovery objective focuses on the SFG $\mathbf{G}$ in frequency-space domain rather than $\tilde{\tilde{\mathbf{G}}}$.

The coherence of a dictionary measures the similarity between its columns~\cite{tropp2004greed}.
Since computing the restricted isometry constants of a given matrix is an NP-hard problem, coherence-based methods are proposed to characterize the recovery capabilities of both the recovery algorithm and the measurement matrix~~\cite{donoho2001uncertainty,elad2002generalized}.
The coherence in the general sense is defined as $\mu=\max_{i\neq j}\left|\mathbf{m}_{i}^H \mathbf{m}_j\right|,\forall i,j \in \{1,\ldots,MK\}$. 
However, the nonzero entries do not appear randomly across all possible positions; instead, they exhibit a structured grouping pattern. The  \textit{group-coherence} is therefore defined as the maximum singular value of the correlation matrix between each pair of column groups, i.e., 
\begin{equation}
    \mu_{\mathcal{J}}=\max _{i,j \in \{1,\ldots,MK\}:i \neq j} \frac{1}{\gamma} \rho(\mathbf{M}^H[i]\mathbf{M}[j]),
    \label{group coherence}
\end{equation}
where $\rho(\mathbf{A})=\sqrt{\lambda_{max}(\mathbf{A}^H\mathbf{A})}$, and  recall that $\mathcal{J} \triangleq\left\{\mathcal{I}_j\right\}_{j=1}^J$, $\mu_{B}=\mu$ as $\gamma=1$. The local property characterized by the \textit{sub-coherence} of $\mathbf{M}$ is defined as
\begin{equation}
    \nu=\max _{j \in \{1,\ldots,J\}} \ \max _{n \neq m : \mathbf{m}_n, \mathbf{m}_m \in \mathbf{M}[j]}\left|\mathbf{m}_n^H \mathbf{m}_m\right|.
    \label{subcoherence}
\end{equation}
The aforementioned types of coherence pertain to global and local characteristics, respectively. 
Technically, the support of leakage is an hyperbola-like region in the delay-wavenumber grid. For the ease of computation and grouping, we can express the leakage in the delay domain and wavenumber domain separately by relaxing the support into rectangles.

\begin{lemma}
    The sparsity degradation coefficient corresponds to the count of grid samples within the leakage region. Given the bandwidth $B$, the aperture size $W$, the carrier frequency $\omega_c$, the resolutions of the delay domain $\Delta\tau$, and the wavenumber domain $\Delta\mathbf{k}_{\theta}$, we have
    \begin{equation}
        \gamma\leq\frac{32c}{\Delta\tau\Delta\mathbf{k}_{\theta}TBW(2\omega_c-B)}.
    \end{equation}
    \label{group size}
\end{lemma}
\textit{Proof}: See Appendix~\ref{sec:Proof of Lemma 1 and Th2} for the detailed proof.
\begin{remark}
    The number of paths $L$ of the channel can be regarded as the group sparsity of the observation model in~\eqref{observation model}, which means that $\mathbf{x_0}$ is a group $L$-sparse vector. The size of the group is offered by the sparsity degradation coefficient $\gamma$ as given in Lemma \ref{group size}.
\end{remark}
Lemma~\ref{group size} provides guidance for choosing the group size in algorithms focused on group sparsity recovery. It indicates that increased bandwidth and aperture size notably reduce the sparsity degradation, aligning with our intuition.
If there is no prior knowledge about the group-sparsity structure,  $\mathbf{x_0}$ will be treated as a $\gamma L$-sparse vector. Such that a sufficient condition for perfect recovery using OMP or $\ell_1$ norm minimization is $\gamma L<(\mu^{-1}+1)/2$ where $\mu$ is the conventional coherence of $\mathbf{M}$~\cite{tropp2004greed}. With the group structure, the sufficient condition can be replaced by a weaker one $\gamma L< (\mu_{\mathcal{J}}^{-1}+\gamma-(\gamma-1)\nu\mu_{\mathcal{J}}^{-1})/2$,
where the group-coherence and subcoherence are defined in (\ref{group coherence}) and (\ref{subcoherence})~\cite{eldar2010block}.
\begin{theorem}
    Given the delay spread $\tau_{\text{max}}$ of the channel, the bandwidth $B=\beta\omega_c$,  the aperture size $W=N_{\lambda}\lambda_c$ of FAS, and the size of SFG  $M,K$, then we can respectively obtain two leakage coefficients in the delay domain and wavenumber domain
    \begin{equation}
        \gamma_{\tau}=\frac{4K}{\tau_{\text{max}}B\sqrt{T}},
    \end{equation}
    \begin{equation}
        \gamma_{\mathbf{k}}=\frac{2M}{\pi\sqrt{T}N_{\lambda}(2-\beta)},
    \end{equation}
    where $T$ is the power detection threshold, and $N_\lambda$ denotes the multiple of the wavelength.
    \label{theorem 2}
\end{theorem}
\textit{Proof}: See Appendix~\ref{sec:Proof of Lemma 1 and Th2} for the detailed proof.

Theorem~\ref{theorem 2} further quantifies leakage coefficients in delay and wavenumber domains, linking them to system parameters such as bandwidth $B$, aperture size $W$ and spread delay $\tau_{\text{max}}$.   
$\tau_{\text{max}}B$ denotes the bandwidth-delay product, which is commonly used to characterize the number of independent paths that can be provided in the delay-frequency domain. $N_{\lambda}$ represents the multiplicity of wavelengths that determines the ability to discern separate spatial paths. Higher $\tau_{\text{max}}B$ or $N_{\lambda}$ leads to decreasing of the corresponding sparsity degradations when the resolution is fixed. The power threshold usually depends on the noisy level as
 \begin{equation}
     T =
    \left\{\begin{array}{l}
        \frac{1}{2} \quad, \sigma_{\omega}^2<0.5\\
         \sigma_{\omega}^2 \quad, \sigma_{\omega}^2\geq0.5,
    \end{array}\right. 
    \label{TT}
 \end{equation}
 which is adaptive to noise. As expressed in the Eq.\eqref{T}, the power threshold $T$ defines the support of the leakage. The region satisfying the Eq.\eqref{T} is the leakage support. As established in Eq.\eqref{TT}, the half-power principle (3 dB bandwidth) governs signal reconstruction when compressed observations are noise-free or exhibit high SNR. Conversely, under significant noise variance ($\sigma_{\omega}^2$), we set the threshold $T=\sigma_{\omega}^2$ to exclude noise components from the support region, thereby preserving signal information integrity.

Assuming that $\Delta\omega=\frac{2\pi}{\tau_{\text{max}}}$, we have $\Delta\tau=\frac{2\pi}{B}$, since $\frac{B}{\Delta \omega}=\frac{\tau_{\text{max}}}{\Delta\tau}=K$. Combining with $\operatorname{sinc}(B(\tau_l-\tau)/2)$ in (\ref{leakage pattern}), we have that there is no leakage related to $\tau$. Taking the half-power threshold (i.e., $T=1/2$), we have
\begin{equation}
    \gamma\leq\frac{2\sqrt{2}\omega_c M}{\pi N_{\lambda}(2\omega_c-B)},
    \label{group size regardless of delay's leakage}
\end{equation}
where $W=N_{\lambda}\lambda$. For the narrow band system, (\ref{group size regardless of delay's leakage}) can be simplified as $\gamma\leq \sqrt{2}M/N_{\lambda}\pi$.

To conclude, the analysis in this subsection lays the foundation for robust SFG recovery in scenarios with high matrix coherence and limited observations. It also underscores the importance of group-aware algorithms in mitigating sparsity degradation caused by leakage effects.

\subsection{DC-GOMP for Channel Estimation}
\label{sec:DC-GOMP}
\begin{figure*}[b]
    \centering
    \includegraphics[width=\linewidth]{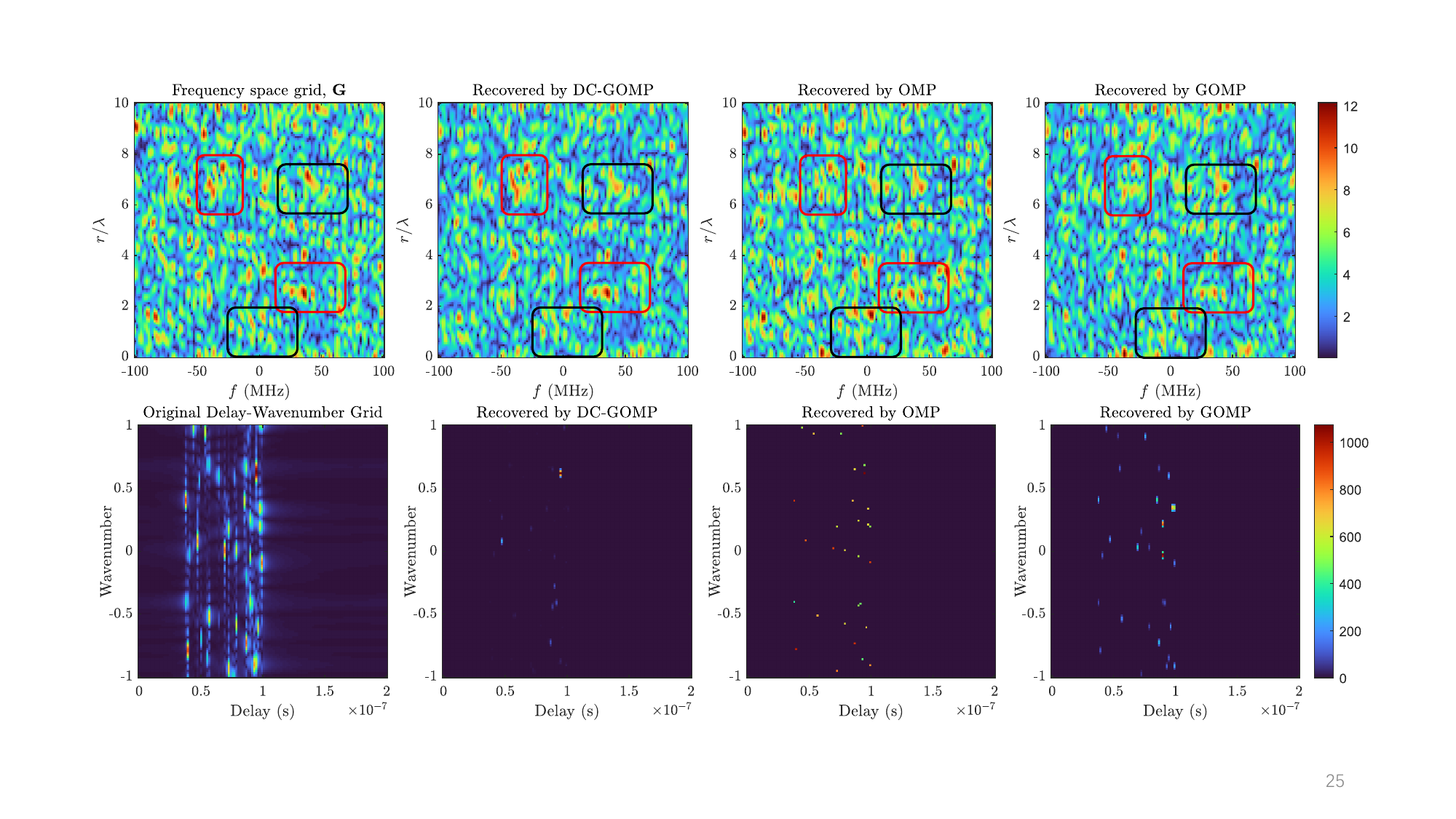}
    \caption{The first row has four expressions in frequency-space domain. The first one represents the original SFG and the last three represent the recovered version by three different algorithm. The second row shows Delay-wavenumber domain expressions corresponding to ones above. Black boxes denote the low power regions and red boxes denote the regions failing to correctly allocate the energy.}
    \label{3algos}
\end{figure*}

As analyzed in Section~\ref{sec:error bound}, traditional sparse recovery algorithms suffer severe performance degradation in frequency-space grid reconstruction due to two inherent limitations: 1) the high redundancy of over-complete dictionaries, and 2) spectral leakage induced by finite aperture size and bandwidth constraints. These limitations are empirically validated by the upper subfigures in Fig.~\ref{3algos}, where conventional methods exhibit significant reconstruction errors.
The orthogonal matching pursuit (OMP) algorithm, while theoretically guaranteeing sparse signal recovery through iterative residual correlation maximization and orthogonal projection updates, proves inadequate for recovering the signal domain expression under redundant dictionary. Specifically, when resolving closely spaced multipath components (common in such environments), OMP tends to over-suppress adjacent paths after selecting a dominant component, as evidenced by missing grid points in Fig.~\ref{3algos}(a). Moreover, its fixed iteration-depth requirement -- mandating prior knowledge of path count -- limits adaptability to varying channel conditions.
To exploit structured sparsity, the group OMP (GOMP) algorithm extends OMP by replacing single-atom selection with group-wise subtraction based on predefined 2D-chunking partitions of the index set $\{1,\dots,MK\}$. This strategy improves leakage effect recovery in delay-wavenumber domains, as shown in Fig.~\ref{3algos}(b)'s tighter support set concentration. However, GOMP's uniform grouping -- adopted due to absent prior path distribution knowledge -- fails to address intra-group energy misallocation caused by high subcoherence $\nu$ (defined in~\eqref{subcoherence}). Consequently, it delivers suboptimal SFG recovery accuracy, as residual energy within erroneously grouped atoms persists through iterations.

The proposed  DC-GOMP algorithm operates by iteratively recovering dominant components of doubly-faded signals in the frequency-space domain. The algorithm initializes with the observed signal $\mathbf{y}_0$, sensing matrix $\mathbf{M}$, sparsifying transform $\mathbf{D}$, and parameters including maximum iteration count $N_{\text{iter}}$, selection cardinality $\gamma = \gamma_\tau \gamma_{\mathbf{k}}$, and residual threshold $\varepsilon_1$. Starting with an empty support set $T_{(0)}$ and initial residual $\mathbf{r}_{(0)} = \mathbf{y}_0$, the iterative process begins. At each iteration $l$, the correlation vector $\mathbf{q} = \mathbf{M}^H \mathbf{r}_{(l-1)}$ is computed to quantify the alignment between the residual and dictionary atoms. The top-$\gamma$ indices $\mathcal{I}_l$ corresponding to the largest magnitudes in $\mathbf{q}$ are selected, breaking away from traditional leakage-constrained grouping to prioritize high-energy bases across delay-wavenumber subspaces. The support set is then updated as $T_{(l)} = T_{(l-1)} \cup \mathcal{I}_l$, and the sparse coefficients $\mathbf{x}_{T_{(l)}}$ are estimated via least-squares minimization using the pseudo-inverse of the submatrix $\mathbf{M}_{T_{(l)}}^\dagger$. The residual $\mathbf{r}_{(l)}$ is recalculated by subtracting the contribution of the estimated components, and the signal $\mathbf{\hat{g}}_{(l)}$ in the fading domain is reconstructed through $\mathbf{D}\mathbf{\hat{x}}_{(l)}$. The process terminates when the residual norm falls below $\varepsilon_1$ or the iteration limit is reached.\footnote{\label{foot: convergence} Focusing on the sparse domain, OMP convergence analysis establishes RIP-based guarantees by characterizing iterative residual norm reduction. In contrast the proposed DC-GOMP shifts its recovery target to signal domain expression which make the conventional analysis unsuitable. Given the inherent mathematical complexity of rigorous convergence rate analysis in the signal domain and manuscript length constraints, the derivation of the convergence rate (as distinct from the error bound) remains an important topic for our future work. The  numerical simulation about the convergence of DC-GOMP, OMP and GOMP are presented in Fig.~\ref{convergence rate}.}

DC-GOMP addresses challenges in doubly-faded signal recovery through three pivotal design principles. First, its dynamic descending-order grouping mechanism replaces static leakage-based group selection with adaptive cross-subspace atom selection, reducing sub-coherence effects and preventing energy misallocation.  By greedily selecting the $\gamma$ most correlated atoms per iteration, DC-GOMP rapidly captures dominant scattering features and converges approximately $\gamma$ times faster than conventional OMP. The time complexity is $\mathcal{O}\left((L/\gamma)M^2K^2\right)$ whereas for OMP it is $\mathcal{O}\left(LM^2K^2\right)$.   $\gamma=\gamma_{\tau}\gamma_{k}$ is an approximation on the number of samples within the support of leakage defined in~\eqref{T} (see Appendix~\ref{sec:Proof of Lemma 1 and Th2}). The selection of gamma governs both the convergence rate and reconstruction precision of the DC-GOMP algorithm. An excessively large gamma value may compromise recovery accuracy, whereas an unduly small gamma impedes convergence speed; thus a dual-effect necessitates rigorous bisection-based calibration. Second, the strategy inherently suppresses both group coherence $\mu_{\mathcal{J}}$ and sub-coherence $\nu$, relaxing the theoretical recovery condition to $\gamma L < \frac{1}{2}(\mu_{\mathcal{J}}^{-1} + \gamma - (\gamma-1)\nu \mu_{\mathcal{J}}^{-1})$~\cite{eldar2010block}. This ensures robustness under practical constraints such as dictionary redundancy and limited aperture/bandwidth-induced leakage. Crucially, DC-GOMP shifts the recovery objective from exact support identification in 2D-sparse domains to direct fidelity optimization in the doubly-fading domain (i.e., space-frequency domain), bypassing the computational complexity of traditional sparse recovery and the mergence of leaky dense paths. The resultant framework achieves a provable balance between reconstruction accuracy and computational efficiency, making it particularly suitable for large-scale systems with high spatial resolution requirements.

\begin{algorithm} 
  \caption{\emph{DC-GOMP}: Descending Correlation Group Orthogonal Matching Pursuit.} 
  \begin{algorithmic}[1]
    \Require $\mathbf{S}, \mathbf{D}, \mathbf{M}, \mathbf{y_{0}}, N_{\text{iter}}, \gamma_{\tau}, \gamma_{\mathbf{k}}, \varepsilon_1$
    \Ensure $\mathbf{\hat{g}}$
    \State Initialize: $l = 0$, $\mathbf{r}_{(0)} = \mathbf{y_0}$, $\gamma=\gamma_{\tau}\gamma_{\mathbf{k}}$, $T_{(0)} = \{\varnothing\}$
    \While{$\|\mathbf{S}\mathbf{\hat{g}}_{(l)} - \mathbf{y_{0}}\|_2 > \varepsilon_1$ \textbf{and} $l\leq N_{\text{iter}}$}
    \State $l=l+1$
    \State $\mathbf{q}=\mathbf{M}^H\mathbf{r}_{(l-1)}$, where $q_{i_n}$ is the $i_n$-th entry of $\mathbf{q}$
    \State sort $\{i_n\}_n^{MK}$ in descending, $q_{i_1}\geq q_{i_2}\cdots\geq q_{i_{MK}}$
    \State $\mathcal{I}_l = \{i_1,i_2,\dots,i_{\gamma}\}$
    \State $T_{(l)}=T_{(l-1)}\cup \mathcal{I}_l$,
    \State $\mathbf{x}_{T_{(l)}}=\mathbf{M}^{\dagger}_{T_{(l)}}\mathbf{y_0}$
    \State $\mathbf{r}_{(l)}=\mathbf{y_0}-\mathbf{M}\mathbf{x}_{T_{(l)}}$
    \State $\mathbf{x}_{T_{(l)}}\xrightarrow{T_{(l)}}\mathbf{\hat{x}}_{(l)}$
    \State $\mathbf{\hat{g}}_{(l)}=\mathbf{D}\mathbf{\hat{x}}_{(l)}$
\EndWhile
\State \Return $\mathbf{\hat{g}}=\mathbf{\hat{g}}_{(l)}$
  \end{algorithmic}
  \label{DCGOMP}
\end{algorithm}

\section{Spatial Equalization\label{section 4}}
After the SFG reconstruction on the receiver side, the receiver then needs to optimize the antenna positions. 
In this section, we reformulate the max-min antenna position optimization of SE in (\ref{position optimation1}) as a MILP problem with guaranteed global optimality by the  branch-and-bound method. To enable the real-time implementation, we reduce the computational complexity by further proposing a GRSIP algorithm, which maintains decent performance in average while reducing the computation time by two orders of magnitude.

The problem in \eqref{position optimation1} can be rewritten as
\begin{subequations}
\label{position optimation2}
\begin{align}       
        \underset{\boldsymbol{\mathcal{I}_s}}{\arg \max}\ \underset{k \in \{1,\ldots,K\}}{\min}
        &\  \|\mathbf{h}_k\|^2_2\\
        \text{s.t.}
        &\ h_{k,n}= G_{i_n,k},\ i_n\in \boldsymbol{\mathcal{I}_s},\\
        &\ \vert\boldsymbol{\mathcal{I}_s}\vert=N_r,\ \boldsymbol{\mathcal{I}_s}\subset \{1,\dots,M\},
\end{align}
\end{subequations}
where $h_{k,n}$ denotes the $n$-th entry of $\mathbf{h}_k$. (\ref{position optimation2}) is a max-min problem with  the objective  to maximize the minimum of the squares of the $\ell_2$ norm of the individual column vectors $\mathbf{h}_k$.  The constraints impose that the set $\boldsymbol{\mathcal{I}_s}$ is of size $N_r$ and each $\mathbf{h}_k$ is a vector of size $N_r$ consisting of the elements in the set of selected row indices
of  the original matrix $\mathbf{G}$. Specifically, the $n$-th element of $\mathbf{h}_k$ is the element of the $k$-th column and the $i_n$-th row of $\mathbf{G}$. 

The index selection problems are typically formulated as integer programming problems. To solve such problems efficiently, it is often necessary to reformulate them as linear or mixed-integer programming models. However, the objective function in this case involves maximizing the minimum value, which presents a challenge due to its segmented and nonlinear nature. In integer programming, decision variables are generally binary, representing whether a specific row index is selected.
Let $ p_i $ be a binary variable for $ i = 1, \dots, M $, where $ p_i = 1 $ indicates that the $i$-th row is selected, and $ p_i = 0 $ indicates that it is not. Since exactly $ N_r $ rows must be selected, the constraint is given by $\sum_{i=1}^{M} p_i = N_r.$ 
Since each vector $ \mathbf{h}_k $ is formed by $k$-th elements of the selected rows of $ \mathbf{G} $
we have
\begin{equation}
\|\mathbf{h}_k\|^2_2 = \sum_{i=1}^{M} p_i \cdot |G_{i,k}|^2.
\end{equation}
 The objective is to maximize the minimum value of $\|\mathbf{h}_k\|^2_2$ over all $k\in \{1,\ldots,K\}$. In other words, the problem is to maximize $ t $, such that for all $ k $, $\sum_{i=1}^{M} p_i \cdot |G_{i,k}|^2 \geq t.$

This  transforms the problem (\ref{position optimation2}) into a mixed-integer linear programming (MILP) problem, where the objective is to maximize $ t $, i.e., 
\begin{subequations}
    \label{linear interger program}
    \begin{align}
        \underset{\mathbf{p}\in\{0,1\}^{M\times 1}}{\operatorname{\arg\min}} &\ -t\\
        \text{s.t.} &\ \mathbf{p}^T \mathbf{g}_k^{\text{abs}} \geq t,\forall k\in\{1,2,\dots,K\}\\
        &\ \|\mathbf{p}\|_0=N_r,\\
        &\  t\geq0,
    \end{align}
\end{subequations}
where the entries of $\mathbf{g}_k^{\text{abs}}\in\mathbb{R}^{M\times1}$ are the entries of $\mathbf{g}_k\in\mathbb{C}^{M\times1}$ in absolute value. With linear constraints, the problem (\ref{position optimation2}) can be formulated as a linear integer programming problem (\ref{linear interger program}).

The key advance of the MILP model is its global optimality guarantee, since 
it can explore all possible combinations via branch-and-bound. Furthermore, by employing upper and lower bound pruning strategies, MILP significantly reduces ineffective search efforts, thereby enhancing computational efficiency and guaranteeing the identification of the global optimality.
However, despite its global optimality, its computational complexity still grows rapidly as the problem size increases with the time complexity $\mathcal{O}(2^M\cdot K)$. Empirical results show that when $M>50$ and $K>100$, the size of the branch-and-bound search tree leads to memory and time costs that exceed practical limits.

\begin{algorithm}
  \caption{\emph{GRSIP}: Greedy Row Selection with Isolated Preselection.} 
  \begin{algorithmic}[1]
    \Require $\mathbf{G}, N_r, N_{\text{init}}, \Delta n$
    \Ensure $\boldsymbol{\mathcal{I}_s}$
    \State Initialize: $l = 0$, $T_{(0)} = \{\varnothing\}$
    \While{$l\leq N_{\text{init}}$}
    \State $i=\underset{i\in T_{(l)}^{c}}{\arg\max}\ \bar{\mathbf{g}}_i,\ \operatorname{s.t.}\ D_1(\{i\},T_{(l)})\leq \Delta n$
    \State $l=l+1$
    \State $T_{(l)}=T_{(l)}\cup \{i\}$
\EndWhile
    \While{$l\leq N_{r}$}
    \State $i=\underset{i\in T_{(l)}^{c}}{\arg\max}\underset{k}{\min}\ \|\mathbf{h}_k\|_2,\ \text{s.t.}\ \mathbf{h}_k\in\text{Col}(\mathbf{H}_{T_{(l)}\cup\{i\}})$
    \State $l=l+1$
    \State $T_{(l)}=T_{(l)}\cup \{i\}$
\EndWhile
\State \Return $\boldsymbol{\mathcal{I}_s}=T_{(l)}$
  \end{algorithmic}
  \label{GRSIP}
\end{algorithm}

Hence, in the following we also propose a greedy algorithm  in Algorithm~$\ref{GRSIP}$ with the time complexity $\mathcal{O}(MKN_r)$. The algorithm starts by choosing $N_{\text{init}}$ positions with the top average channel gains for each subcarrier, maintaining a minimum separation of $\Delta n$ between them, where $D_1(\cdot,\cdot)$ represents the minimal $\ell_1$-norm distance between two point sets. Then it  incrementally adds indices to the candidate set, following the principle of maximizing the minimum subcarrier gain, until $N_r$ positions are chosen. Note that $\mathbf{H}_T$ is formed by the row vectors of $\mathbf{G}$ associated with row indices in $T$.

\begin{remark}[Applications to Other Models]
The theoretical analysis and     algorithms of the two-step framework for  FAS  proposed in this work  could be extended to other problems besides FAS. 
First, the proposed two-step framework could be directly extended to the antenna selection problem with discrete positions, regardless of whether the exact antennas deployment~\cite{wang2020sparse,gao2017massive,mendoncca2019antenna,huang2018wideband}. The group-sparse recovery formulation and D-GRIP analysis can be directly adapted to delay-Doppler domain channel estimation~\cite{rasheed2020sparse,han2016two}, where structures induce similar group-wise sparsity patterns in reconstruction. The DC-GOMP algorithm employs a correlation-aware selection mechanism to dynamically resolve coherence conflicts, offering a systematic and efficient approach to sparse event detection. Then, MILP-based spatial equalization offers new insights for the resource-constrained optimization in RIS
configuration on discrete phase~\cite{xiong2024optimal}. 
These potential extensions highlight that our methodology effectively tackles the unified challenge of sparsity-aware optimization under structured constraints, making it applicable to a wide range of domains, including computational sensing, adaptive control, and beyond.
\end{remark}
\section{Simulation Results}
In this section, we present the performance of the proposed group-sparsity based frequency-space channel estimation algorithm, i.e., DC-GOMP, in comparison to two traditional algorithms (OMP, GOMP), under FAS-assisted wideband SIMO system. The proposed positions optimization methods, i.e., MILP and GRSIP, are also evaluated through the physical layer simulations and in terms of BER.
\subsection{Simulation Setup}
In the simulations, we consider the FAS-assisted SIMO system where the receiver is equipped with the fluid array of different aperture sizes.
We consider the maximal spread delay of the channel as $2\times10^{-7}$ s, the signal bandwidth is $200$ MHz, and the carrier frequency is $5.8$ GHz. To obtain the SG of the doubly-selective frequency-space channel, the SFG with a $128\times128$ grid is employed for the demand of sparsity.
The effectiveness of the SE is evaluated by applying the antenna position optimization together with the MRC and frequency domain equalization at the receiver side.

\subsection{Results of Channel Estimation}
We demonstrate the recovery performance of the SFG and its corresponding delay-wavenumber domain in Fig.~\ref{3algos} by using OMP, GOMP, and the proposed DC-GOMP scheme. Cool-toned hues denote lower magnitude values, whereas warm-toned hues indicate higher magnitude values (subject to data normalization within the designated scale). Here, we consider $M=K=128$, $N_r=20$, $N_p=40$, $L=40$, where the SNR is set to $10$ dB and the aperture size is $10$ times the wavelength. Since the actual number of paths is not known at the receiver, we consider $50$ iterations for all the algorithms and plot the performance corresponding to the best results for each algorithm. 
Obviously, our proposed DC-GOMP outperforms the other two in the details of the low power region (i.e., black boxes). Fig. \ref{3algos} shows DC-GOMP outperforming benchmark algorithms in signal domain (frequency-space) recovery despite imperfect sparse domain (delay-wavenumber) recovery, which confirms our analysis about the coherent dictionary in Section~\ref{section III}. The energy allocation within the red box of OMP and GOMP mismatches, but DC-GOMP does not. The relative error is computed by
\begin{equation}
    \text{Relative Error}=\|\mathbf{\hat{G}}/\|\mathbf{\hat{G}}\|_F-\mathbf{G}/\|\mathbf{G}\|_F\|_F,
    \label{relative error}
\end{equation}
where the three algorithms (DC-GOMP, OMP, GOMP) respectively have relative errors of $0.24975$, $0.33675$ and $0.36459$. 
Additionally, the initial antennas' positions and the pilots' positions do have some impact on the performance of the recovery. In our simulations, the channel parameters are generated according to the uniformly stochastic distributions described in (\ref{wireless P2P channel}), which leads to better recovery performance for uniformly distributed initial positions.
\begin{figure}[htbp]
    \centering
    \includegraphics[width=0.9\linewidth]{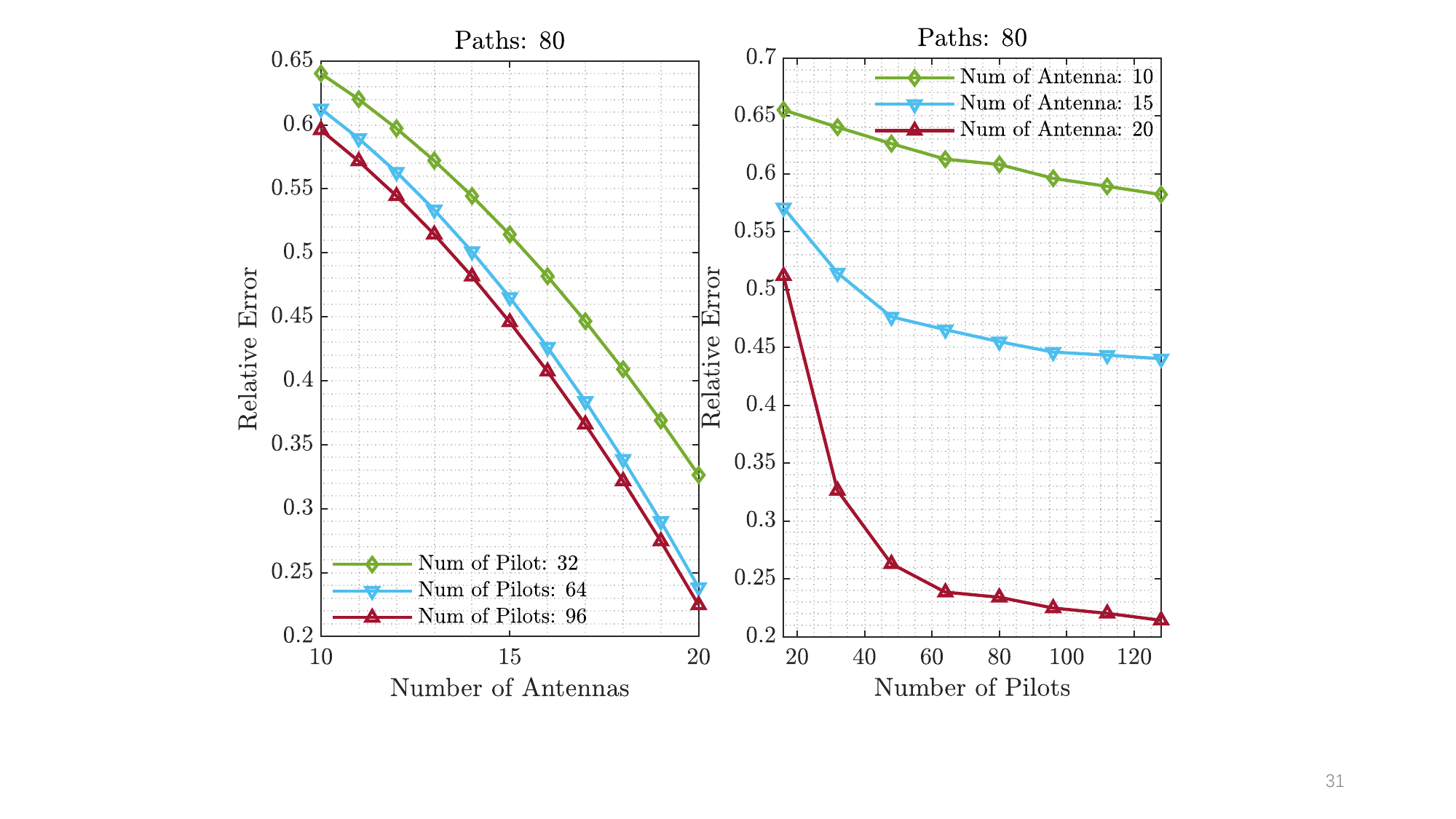}
    \caption{The performance trends of relative error of recovering SFG as the numbers of pilots and antennas varying. }
    \label{23}
\end{figure}

In Fig.~\ref{23}, we evaluate the performance scaling law with respect to the number of antennas and pilots. We consider an aperture size of $10\lambda$ and SNR of $10$ dB, where the number of paths is $80$. Due to the assumption of one time slot, the number of antennas denotes the number of spatial samples, while the number of pilots denotes the number of spectral samples. Obviously, the number of spatial samples affects the relative error of the recovery more than the number of pilots. According to Nyquist's sampling theorem, spatial feature recovery for an aperture of $10\lambda$ requires at least $20$ spatial sampling points. Nonetheless, given the prior knowledge of the physical modeling of propagation and the assumption of group sparsity, a sample of fewer than $20$ antennas is sufficient for effective channel recovery. Generally, most SFG details can be successfully recovered when the relative error is less than $0.5$, e.g., the relative error of recovered SFG in the first row and second column of Fig.~\ref{3algos} is $0.3768$. As demonstrated in the right side of the graph above, with $20$ antennas, the gain from increasing the number of guides decreases significantly beyond $40$ in one-time slot recovery. By maintaining a constant channel over several time slots, multiple observations allow for a significant reduction in terms of the pilot overhead.
\begin{figure}[htbp]
    \centering
    \includegraphics[width=0.9\linewidth]{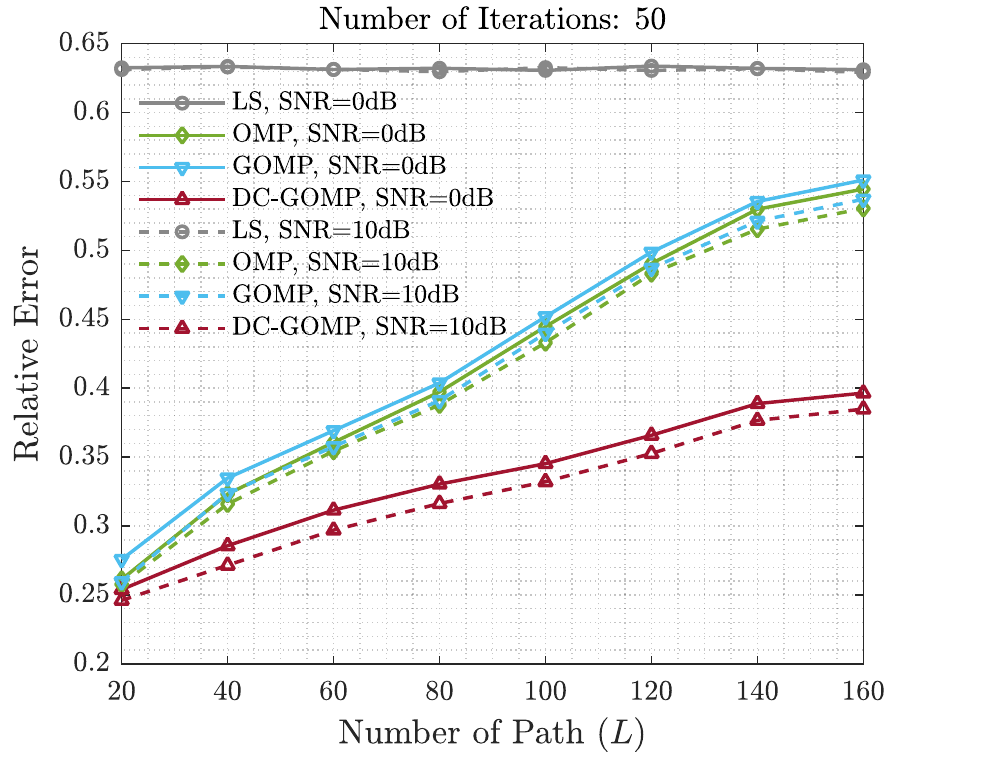}
    \caption{Comparison of four different algorithms on the performance of SFG recovering versus the number of paths.}
    \label{55}
\end{figure}

\begin{figure}[htbp]
    \centering
    \includegraphics[width=0.9\linewidth]{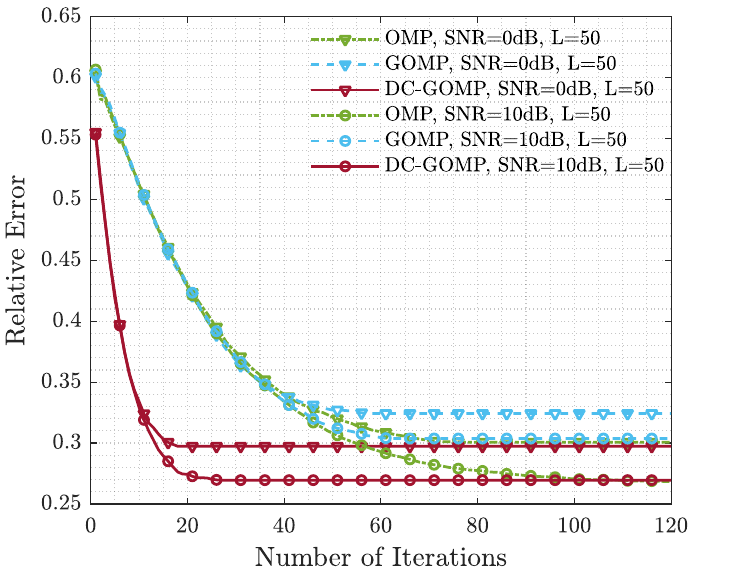}
    \includegraphics[width=0.9\linewidth]{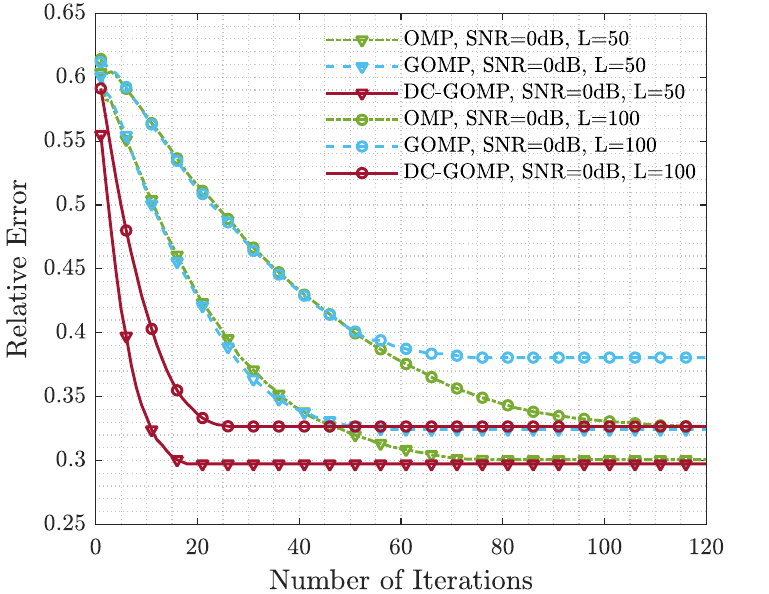}
    \caption{Convergence rate of three algorithms for different SNRs and numbers of paths. }
    \label{convergence rate}
\end{figure}

Fig.~\ref{55} and Fig.~\ref{convergence rate} respectively show the superiority of our proposed DC-GOMP on the performance and the convergence rate compared to the classical OMP and GOMP. All the results in Fig.~\ref{55} are evaluated under the $50$ iterations. As the number of paths rises, DC-GOMP significantly outperforms OMP and GOMP, while traditional methods like least square (LS) fail to work due to the fact that they cannot exploit the sparsity. For environments with large numbers of paths, the leakage effects of these paths hinder accurate recovery in the sparse domain as analyzed in Section \ref{section 3A}, which significantly degrades the capability of traditional sparse recovery methods.
Fig.~\ref{convergence rate} shows the convergence performances of DC-GOMP, OMP and GOMP. For each iteration of all three matching pursuit-like algorithms, the primary computational demands are associated with the LS and residual projection, i.e., step 8 and step 4 in algorithm \ref{DCGOMP}. That means the time complexity of DC-GOMP is much less than that of the other two. According to the analysis in Section~\ref{sec:DC-GOMP}, the DC-GOMP algorithm converges around $8$ times faster than the conventional OMP algorithm as the $\gamma=8$ here.

\subsection{Results of Spatial Equalization}
\begin{table}
\centering
\caption{Comparison the computation time of three algorithms when $M=K=128$, $N_r=20$ }
\begin{tabular}{ccc}
\hline
\textbf{Algorithms} & \textbf{Average time(s)} & \textbf{Time Complexity} \\ \hline
Branch-and-Bound                  & 90                       & $\mathcal{O}(2^MK)$     \\
GRSIP               & 0.003                    & $\mathcal{O}(MKN_r)$                       \\
Random100           & 0.0025                   & $\mathcal{O}(KN_r)$       \\\hline             
\end{tabular}
\label{tab}
\end{table}
As noted in Section~\ref{section 4}, the computational complexities of both the branch-and-bound (BB) approach and the proposed GRSIP method are analyzed in parallel, with respective orders of $\mathcal{O}(2^M\cdot K)$ and $\mathcal{O}(MKN_r)$. Table~\ref{tab} summarizes the average computation times under specific parameters, with our computer (MATLAB R2024b, i7-11700K).  
We record the computational times of the BB algorithm, the proposed GRSIP method, and (i.e., the random traversal schemes with $100$ and $20$ position combinations respectively, then selecting the optimal outcome),  which are 90s, 0.003s, and 0.0025s  respectively, for the case of $K=128$ and $M=128$. It can be seen that the proposed GRSIP method significantly reduces the complexity of the BB algorithm.

\begin{figure*}[htbp]
    \centering
    \includegraphics[width=\linewidth]{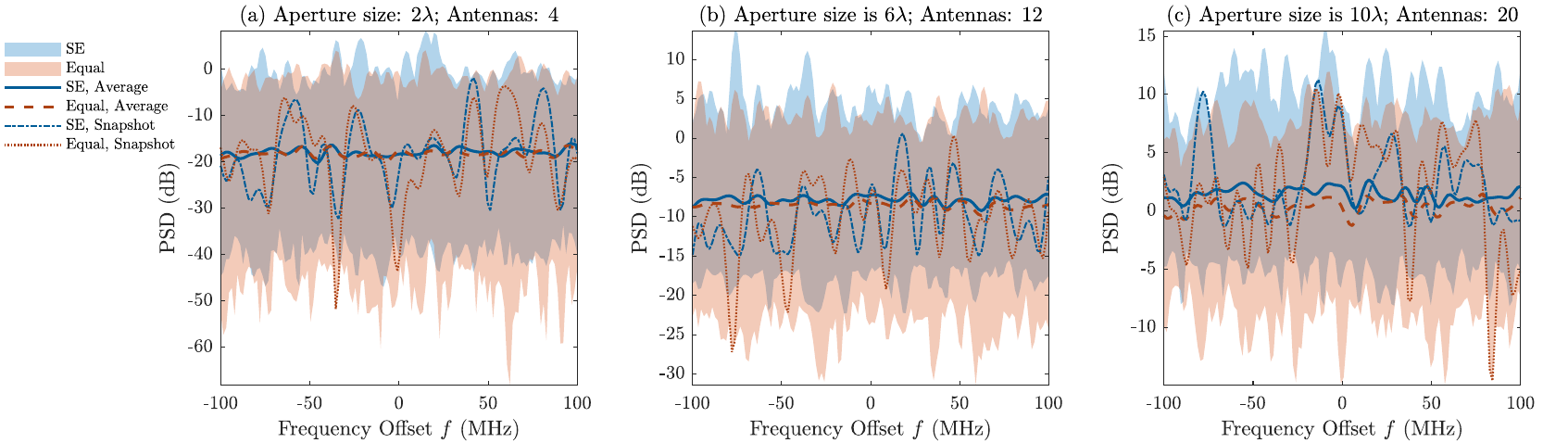}
    \caption{ The combined PSDs demonstrate the application of SE by FAS versus no SE by a fixed equally spaced array for different aperture sizes. The colored shadows and bolder lines respectively denote the range and the average of combined PSDs of 100 times trials. The thinner line denotes one snapshot of two configurations.}
    \label{SE demo}
\end{figure*}

Fig.~\ref{SE demo} demonstrates the role of the FAS when performing SE under different aperture sizes ($2\lambda$, $6\lambda$, and $10\lambda$), corresponding to different numbers of antennas. The term `Equal' refers to the maximum ratio combined channel response of the fixed equal-spaced array, and the term `SE' refers to that of the spatially equalized FAS. Due to the global optimality guarantee, the result of SE in Fig.~\ref{SE demo} is obtained by solving the MILP by the branch-and-bound method. The distinct coverage regions, represented by the blue and orange shaded areas, visually demonstrate the effectiveness of SE. Notably, the worst gain of the subcarrier of the blue region exhibits an improvement of 10$\sim$15 dB compared to the orange region, confirming SE's capability to significantly reduce the probability of deep fading events. We find that SE's effectiveness increases with larger aperture sizes. The SE not only reduces deep fading but also boosts the average SNR of the combined channel by around 2 dB in the case of 10$\lambda$, as indicated by `average' in Fig.~\ref{SE demo}. This average SNR improvement scales with aperture size.

\begin{figure}[htbp]
    \centering
    \includegraphics[width=0.92\linewidth]{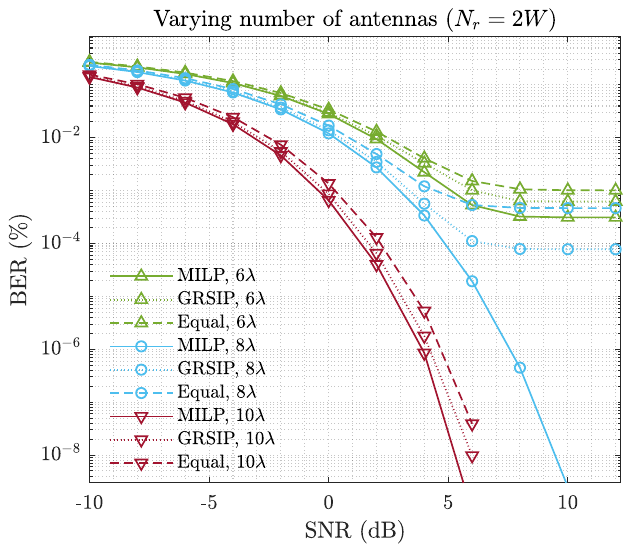}
    \caption{{Comparison of the BER for different aperture size configurations under three position arrangement strategy versus SNR.}}
    \label{BER}
\end{figure}
Fig.~\ref{BER} demonstrates the BER performance for different position arrangements, including equal-spaced, MILP, and GRSIP. The simulation setup includes aperture sizes of $6\lambda$, $8\lambda$, and $10\lambda$, corresponding to 12, 16, and 20 antennas, respectively.
Under the QPSK constellation, we observe that the other two SE algorithms outperform the equal-spaced arrangement in terms of BER, especially with high SNR. 
We compare the $6\lambda$ aperture and 12 antennas FAS-assisted receiver applying SE to the `Equal', where the `MILP' eliminates the error floor up to a BER of $10^{-8}$, and the `GRSIP' lowers the error floor below a BER of $10^{-4}$. It is observed that "MILP, 6$\lambda$" outperforms "Equal, 8$\lambda$", indicating that SE can achieve comparable performance with fewer antennas or smaller aperture sizes.

\begin{figure}[htbp]
    \centering
    \includegraphics[width=0.92\linewidth]{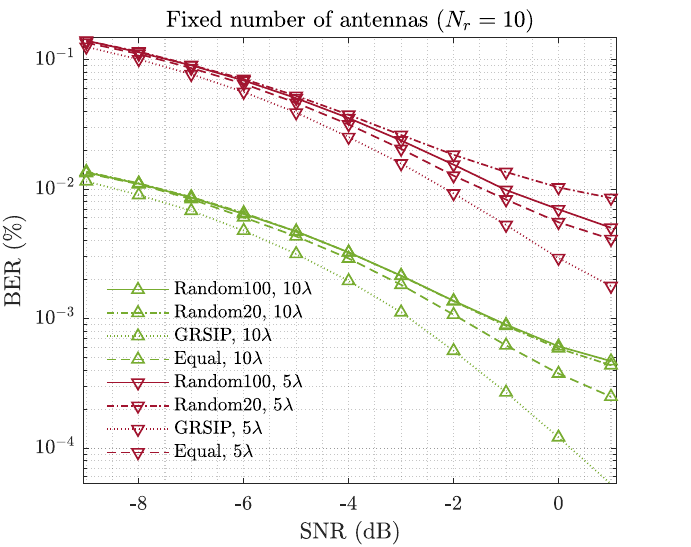}
    \caption{ Comparison of the BER for different aperture size configurations under Four position arrangement strategies versus SNR with fixed number of antennas.}
    \label{BER2}
\end{figure}
\begin{figure}[htbp]
    \centering
    \includegraphics[width=0.92\linewidth]{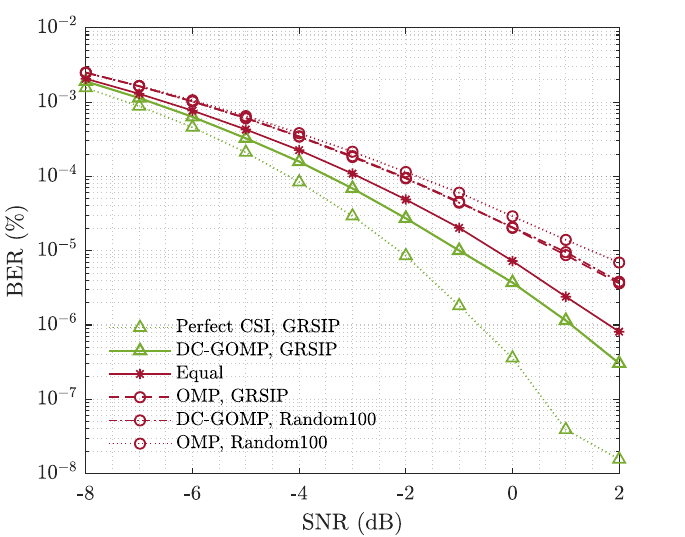}
    \caption{ Comparison of the BER for different joint channel estimation and spatial equalization strategies.}
    \label{BER3}
\end{figure}
Fig.~\ref{BER2} shows the performance advantage of the GRSIP algorithm over other low-complexity methods. `Random100' and `Random20' denote the random traversal schemes with $100$ and $20$ position combinations, respectively, selecting the optimal outcome. Unlike the narrowband FAS, these approaches suffer from substantial degradation, because deep-fading frequencies exhibit spatial coupling, not independence. Consequently, relocating the antenna within a limited area shifts deep-fading frequencies across the bandwidth without eliminating them. Thus, simple randomized antenna placement fails to achieve spatial diversity gains for spatial equalization. Instead, when two antennas are positioned too closely, it actually degrades diversity gain due to mutual coupling effects. Fig.~\ref{BER3} illustrates the performance when channel estimation and spatial equalization operate jointly. It demonstrates that under identical channel estimates obtained via the DC-GOMP algorithm, the GRSIP approach achieves significantly superior results compared to the random selection method, and outperforms the fixed uniform-spacing arrays. This underscores that the accuracy of channel reconstruction profoundly impacts the continuum of performance advantages achievable through position optimization in wideband FAS, particularly those reliant on precise spatial configuration.

\section{Conclusion}
This work established a unified framework for channel estimation and spatial equalization in FAS. We proposed a group-sparse recovery method with theoretical guarantees to address sparsity degradation under practical aperture constraints. A dynamic correlation-aware algorithm (DC-GOMP) was developed to overcome coherence limitations in sparse reconstruction, demonstrating significant improvements over conventional compressive sensing methods. Furthermore, the spatial equalization formulation (i.e., MILP) bridged combinatorial configurations of antennas' positions  with continuous channel characteristics, enabling FAS to achieve performance comparable to larger fixed arrays. The framework establishes a foundational connection between wireless propagation and computational signal processing, advancing FAS implementations for 6G communications. Future work contains (i) the  extension of  the proposed scheme to dynamic multi-user environments and learning-based channel adaptation, specifically investigating robust channel estimation techniques to mitigate multi-user interference and leverage spatial-temporal correlations in FAS deployments; (ii) and the implementation of the proposed scheme into a practical FAS prototype.

\appendices
\section{Proof of Theorem~\ref{Error Bound}}
\label{sec:append proof of th1}
\subsection{Preliminary}
Rewrite the noisy observation model as $\mathbf{y_0}=\mathbf{S}\mathbf{g_0}+\mathbf{z}=\mathbf{S}\mathbf{D}\mathbf{x_0}+\mathbf{z}$ and the modified convex second order cone program,
\begin{equation}
\label{P2}
\begin{aligned}
    \mathbf{\hat{g}}=\underset{\mathbf{g}}{\arg \min}
    &\left\|\mathbf{D}^H\mathbf{g}\right\|_{2,\mathcal{J}}\\
    \text{s.t.}&\left\| \mathbf{S}\mathbf{g} - \mathbf{y_0} \right\|_2\leq\varepsilon.
\end{aligned}
\end{equation}
Denoting the $(\mathbf{D}^H\mathbf{g})^{(k)}$ as the vector consisting of the largest $k$ groups in $\mathbf{D}^H\mathbf{g}$.
Assuming $\mathbf{x_0}$ is a group $k$-sparse vector and $\mathbf{SD}$ satisfies the group RIP with restricted isometry constants $\delta_k<\sqrt{2}-1$, the group RIP is defined in~\cite{eldar2009robust}.

Denoting by $\mathbf{D}_{T}$ the matrix $D$ restricted to the column-groups indexed by $T$, which means that $\mathbf{D}_{T}$ is the same as $D$ only at the columns with indices $i\in T$ where $T$ is the union of certain column-groups.
Let $T_0$ denote the set of the largest $k$ group of $\mathbf{D}^H\mathbf{g}$ in $\ell_{2,\mathcal{J}}$-norm. Divide the groups' indices $T_0^c$ into a set of size $M$ in order to decrease $\ell_{2,\mathcal{J}}$-norm of $\mathbf{D}^H\mathbf{h}$ where $\mathbf{h}=\mathbf{g_0}-\mathbf{\hat{g}}$, i.e., $\|\mathbf{D}_{T_1}^H\mathbf{h}\|_{2,\mathcal{J}}\geq\|\mathbf{D}_{T_2}^H\mathbf{h}\|_{2,\mathcal{J}}\geq\cdots$. $P$ is usually taken as an integer multiple of $k$, and $P>k$.

Two common inequalities are given here, for any values $u$,$v$ and $c>0$,
\begin{equation}
    uv\leq\frac{cu^2}{2}+\frac{v^2}{2c},\ \sqrt{u^2+v^2}\leq u+v.
    \label{lemma 2.6}
\end{equation}
\subsection{Bounding the Tail} The tail of recovery can be expressed as $\sum_{i\geq2}\|\mathbf{D}^H_{T_i}\mathbf{h}\|_{2}$.
Since the $\mathbf{\hat{y}}$ is the global optimum of problem (\ref{P2}) so that
\begin{equation*}
    \begin{aligned}
        &\|\mathbf{D}^H\mathbf{g_0}\|_{2,\mathcal{J}}\\
        =&\|\mathbf{D}^H_{T_0}\mathbf{g_0}\|_{2,\mathcal{J}}+\|\mathbf{D}^H_{T_0^c}\mathbf{g_0}\|_{2,\mathcal{J}}\\
        \geq &\|\mathbf{D}^H_{T_0}(\mathbf{g_0-h})\|_{2,\mathcal{J}}+\|\mathbf{D}^H_{T_0^c}(\mathbf{g_0-h})\|_{2,\mathcal{J}}\\
        \geq  &\|\mathbf{D}^H_{T_0}\mathbf{g_0}\|_{2,\mathcal{J}}-\|\mathbf{D}^H_{T_0}\mathbf{h}\|_{2,\mathcal{J}}+\|\mathbf{D}^H_{T_0^c}\mathbf{h}\|_{2,\mathcal{J}}-\|\mathbf{D}^H_{T_0^c}\mathbf{g_0}\|_{2,\mathcal{J}},
    \end{aligned}
\end{equation*}
which leads to the cone constraint 
\begin{equation}
    \|\mathbf{D}_{T_0^c}^H\mathbf{h}\|_{2,\mathcal{J}}\leq 2\|\mathbf{D}_{T_0^c}^H\mathbf{g_0}\|_{2,\mathcal{J}}+\|\mathbf{D}_{T_0}^H\mathbf{h}\|_{2,\mathcal{J}}
    \label{cone constraint}
\end{equation}
For simplicity of notation set $T_{01}=T_0\bigcup T_1$, $\rho=k/P$ and the tail as $\eta=2\|\mathbf{D}_{T_0^c}^H\mathbf{g_0}\|_{2,\mathcal{J}}$. We have
\begin{equation}
    P^{\frac{1}{2}}\sum_{i\geq2}\|\mathbf{D}^H_{T_i}\mathbf{h}\|_{2}\leq \sum_{i\geq1}\|\mathbf{D}^H_{T_i}\mathbf{h}\|_{2,\mathcal{J}}=\|\mathbf{D}^H_{T_0^c}\mathbf{h}\|_{2,\mathcal{J}}
\end{equation}
and Cauchy-Schwarz leads to
\begin{equation}
    \|\mathbf{D}^H_{T_0}\mathbf{h}\|_{2,\mathcal{J}}\leq k^{\frac{1}{2}} \|\mathbf{D}^H_{T_0}\mathbf{h}\|_{2}.
\end{equation}
Then along with cone constraint in (\ref{cone constraint}) we have
\begin{equation}
   \sum_{i\geq2}\|\mathbf{D}^H_{T_i}\mathbf{h}\|_{2}\leq \sqrt{\rho}\left( 
\|\mathbf{D}_{T_0}^H\mathbf{h}\|_2+\eta \right)
\label{lemma2.2}
\end{equation}
\subsection{Consequence of D-GRIP}
The observation error vector $\mathbf{Sh}$ satisfies the following
\begin{equation}
\begin{aligned}
    \|\mathbf{Sh}\|_2&=\|\mathbf{Sg_0}-\mathbf{S\hat{g}}\|_2\\
    &\leq\|\mathbf{Sg_0}-\mathbf{y_0}\|_2+\|\mathbf{S\hat{g}}-\mathbf{y_0}\|_2
    \leq 2\varepsilon.
\end{aligned}
\label{noisy error}
\end{equation}
Since $\mathbf{D}$ is a tight frame and normalized by $\mathbf{D}\mathbf{D}^H$,  along with D-GRIP and (\ref{lemma2.2}), we have
\begin{equation}
    \begin{aligned}
        &\|\mathbf{Sh}\|_2 \\
        =&\|\mathbf{S}\mathbf{D}\mathbf{D}^H\mathbf{h}\|_2\\
        \geq& \|\mathbf{S}\mathbf{D}_{T_{01}}\mathbf{D}_{T_{01}}^H\mathbf{h}\|_2-\sum_{i\geq 2}\|\mathbf{S}\mathbf{D}_{T_{i}}\mathbf{D}_{T_{i}}^H\mathbf{h}\|_2\\
        \stackrel{\text{D-GRIP}}{\geq}& \sqrt{1-\delta_{k+P}}\|\mathbf{D}^H_{T_{01}}\mathbf{h}\|_2-\sqrt{1+\delta_{P}}\sum_{i\geq 2}\|\mathbf{D}^H_{T_{i}}\mathbf{h}\|_2\\
        \stackrel{(\ref{lemma2.2})}{\geq}& \sqrt{1-\delta_{k+P}}\|\mathbf{D}^H_{T_{01}}\mathbf{h}\|_2-\sqrt{\rho(1+\delta_{P})}\left( 
        \|\mathbf{D}_{T_0}^H\mathbf{h}\|_2+\eta \right)\\
        \stackrel{(\ref{lemma2.2})}{\geq}& \sqrt{1-\delta_{k+P}}\|\mathbf{D}^H_{T_{01}}\mathbf{h}\|_2-\sqrt{\rho(1+\delta_{P})}\left( 
    \|\mathbf{h}\|_2+\eta \right)
    \end{aligned}
    \label{consequence of GRIP}
\end{equation}
Combining (\ref{noisy error}) and (\ref{consequence of GRIP}), we have
\begin{equation}
    \sqrt{1-\delta_{k+P}}\|\mathbf{D}^H_{T_{01}}\mathbf{h}\|_2-\sqrt{\rho(1+\delta_{P})}\left( 
    \|\mathbf{h}\|_2+\eta \right)\leq 2\varepsilon
    \label{lemma 2.4}
\end{equation}
\subsection{Bounding the Error}
The error vector $\mathbf{h}$ has the norm that satisifies
\begin{equation}
    \begin{aligned}
        &\|\mathbf{h}\|_2^2
        =\|\mathbf{D}^H\mathbf{h}\|_2^2\\
        \leq&\|\mathbf{D}^H_{T_{01}}\mathbf{h}\|_2^2+\|\mathbf{D}^H_{T_{01}^c}\mathbf{h}\|_2^2\\
        \stackrel{(\ref{lemma2.2})}{\leq}&\|\mathbf{D}^H_{T_{01}}\mathbf{h}\|_2^2+\rho\left( \|\mathbf{h}\|_2 +\eta\right)^2\\
        \leq& \|\mathbf{D}^H_{T_{01}}\mathbf{h}\|_2^2+\rho\left(\|\mathbf{h}\|_2^2+2\eta\|\mathbf{h}\|_2+\eta^2\right)\\
        \stackrel{(\ref{lemma 2.6})}{\leq}&\|\mathbf{D}^H_{T_{01}}\mathbf{h}\|_2^2+\rho\left(\|\mathbf{h}\|_2^2+c\|\mathbf{h}\|^2_2+\eta^2/c++\eta^2\right)\\
        \stackrel{(\ref{lemma 2.6})}{\leq}&\frac{1}{\sqrt{1-\rho-\rho c}}
        \left(\|\mathbf{D}^H_{T_{01}}\mathbf{h}\|_2+\eta\sqrt{\rho(1+1/c)} \right)\\
        \stackrel{(\ref{lemma 2.4})}{\leq}&\frac{2\varepsilon+\sqrt{\rho(1+\delta_P)(\|\mathbf{h}\|_2+\eta)}}{\sqrt{1-\delta_{k+P}}\sqrt{1-\rho-\rho c}}+\frac{\eta\sqrt{\rho(1+1/c)}}{\sqrt{1-\rho-\rho c}}
    \end{aligned}
    \label{lemma 2.5}
\end{equation}
Substituting $\eta$ into (\ref{lemma 2.5}), we have the conclusion
\begin{equation}
    \|\mathbf{h}\|_2\leq C_0\|\mathbf{D}^H\mathbf{g_0}-(\mathbf{D}^H\mathbf{g_0})^{(k)}\|_{2,\mathcal{J}}+C_1 \varepsilon
\end{equation}
where 
\begin{equation}
    C_0=2\frac{a\sqrt{1-\delta_{k+P}}+\sqrt{1+\delta_P}}{b\sqrt{1-\delta_{k+P}}-\sqrt{1+\delta_P}},
\end{equation}
\begin{equation}
    \ C_1=\frac{2}{b\sqrt{\rho}\sqrt{1-\delta_{k+P}}-\sqrt{1+\delta_P}},
\end{equation}
and $c$ is any positive value, $a=\sqrt{1+\frac{1}{c}},\ b=\sqrt{\frac{1}{\rho}-1-c}.$

$\hfill\blacksquare$ 
\section{Proof of Lemma~\ref{group size} and Theorem~\ref{theorem 2}}
\label{sec:Proof of Lemma 1 and Th2}
The hyperbola-like support $(\theta,\tau)$ of the leakage satisfies that
\begin{equation}
    \left(\frac{8 c}{W B\left(2 \omega_c-B\right)\left|\cos \theta_l-\cos \theta\right|\left|\tau_l-\tau\right|}\right)^2\geq T,
    \label{T}
\end{equation}
which can be relaxed into rectangle-like region as
\begin{align}
   & \left(\frac{4 c}{W\left(2 \omega_c-B\right)\left|\cos \theta_l-\cos \theta\right|}\right)^2\geq T,
    \label{angular leakage}\\
&     \left(\frac{2}{B\left|\tau_l-\tau\right|}\right)^2\geq T.
\end{align}
In order to guarantee the size of leakage support at different angle, we transform the wavenumber domain into the wavenumber domain, i.e., $\mathbf{k}_{\theta}=\cos\theta$, so that (\ref{angular leakage}) becomes
\begin{equation}
    \left(\frac{4 c}{W\left(2 \omega_c-B\right)\left|\mathbf{k}_{\theta_l}-\mathbf{k}_{\theta}\right|}\right)^2\geq T.
    \label{wavenumber leakage}
\end{equation}
By simple arithmetic, we have
\begin{equation}
    \frac{2}{B\sqrt{T}}\geq \vert\tau_l-\tau\vert,\ \frac{4c}{W\sqrt{T}(2\omega_c-B)}\geq\vert\mathbf{k}_{\theta_l}-\mathbf{k}_{\theta}\vert.
    \label{lemma 1 proof}
\end{equation}
so that given the resolutions in delay domain and wavenumber domain, Lemma~\ref{group size} can be obtained. 

Using the identities $    \frac{B}{\Delta\omega}=\frac{\tau_{\text{max}}}{\Delta\tau}=K, \frac{2}{\Delta\mathbf{k_{\theta}}}=\frac{W}{\Delta r}=M,$
and (\ref{lemma 1 proof}), Theorem~\ref{theorem 2} can be obtained.

$\hfill\blacksquare$
\ifCLASSOPTIONcaptionsoff
  \newpage
\fi

\bibliographystyle{IEEEtran}
\bibliography{ref}

\begin{thebibliography}{10}
\providecommand{\url}[1]{#1}
\csname url@samestyle\endcsname
\providecommand{\newblock}{\relax}
\providecommand{\bibinfo}[2]{#2}
\providecommand{\BIBentrySTDinterwordspacing}{\spaceskip=0pt\relax}
\providecommand{\BIBentryALTinterwordstretchfactor}{4}
\providecommand{\BIBentryALTinterwordspacing}{\spaceskip=\fontdimen2\font plus
\BIBentryALTinterwordstretchfactor\fontdimen3\font minus \fontdimen4\font\relax}
\providecommand{\BIBforeignlanguage}[2]{{%
\expandafter\ifx\csname l@#1\endcsname\relax
\typeout{** WARNING: IEEEtran.bst: No hyphenation pattern has been}%
\typeout{** loaded for the language `#1'. Using the pattern for}%
\typeout{** the default language instead.}%
\else
\language=\csname l@#1\endcsname
\fi
#2}}
\providecommand{\BIBdecl}{\relax}
\BIBdecl

\bibitem{wong2020fluid}
K.-K. Wong, A.~Shojaeifard, K.-F. Tong, and Y.~Zhang, ``Fluid antenna systems,'' \emph{IEEE Trans. Wireless Commun.}, vol.~20, no.~3, pp. 1950--1962, 2020.

\bibitem{wong2021fluid}
K.-K. Wong and K.-F. Tong, ``Fluid antenna multiple access,'' \emph{IEEE Trans. Wireless Commun.}, vol.~21, no.~7, pp. 4801--4815, 2021.

\bibitem{wong2022bruce}
K.-K. Wong, K.-F. Tong, Y.~Shen, Y.~Chen, and Y.~Zhang, ``Bruce lee-inspired fluid antenna system: Six research topics and the potentials for 6g,'' \emph{Front. Commun. Networks}, vol.~3, p. 853416, 2022.

\bibitem{wong2020performance}
K.~K. Wong, A.~Shojaeifard, K.-F. Tong, and Y.~Zhang, ``Performance limits of fluid antenna systems,'' \emph{IEEE Commun. Lett.}, vol.~24, no.~11, pp. 2469--2472, 2020.

\bibitem{zhu2024historical}
L.~Zhu and K.-K. Wong, ``Historical review of fluid antenna and movable antenna,'' \emph{arXiv preprint arXiv:2401.02362}, 2024.

\bibitem{huang2021liquid}
Y.~Huang, L.~Xing, C.~Song, S.~Wang, and F.~Elhouni, ``Liquid antennas: Past, present and future,'' \emph{IEEE Open Journal of Antennas and Propagation}, vol.~2, pp. 473--487, 2021.

\bibitem{zhang2024novel}
J.~Zhang, J.~Rao, Z.~Li, Z.~Ming, C.-Y. Chiu, K.-K. Wong, K.-F. Tong, and R.~Murch, ``A novel pixel-based reconfigurable antenna applied in fluid antenna systems with high switching speed,'' \emph{IEEE Open Journal of Antennas and Propagation}, 2024.

\bibitem{hoang2021computational}
T.~V. Hoang, V.~Fusco, T.~Fromenteze, and O.~Yurduseven, ``Computational polarimetric imaging using two-dimensional dynamic metasurface apertures,'' \emph{IEEE Open Journal of Antennas and Propagation}, vol.~2, pp. 488--497, 2021.

\bibitem{wang2025electromagnetically}
R.~Wang, P.~Zheng, V.~V. Kotte, S.~Rauf, Y.~Yang, M.~M.~U. Rahman, T.~Y. Al-Naffouri, and A.~Shamim, ``Electromagnetically reconfigurable fluid antenna system for wireless communications: Design, modeling, algorithm, fabrication, and experiment,'' \emph{arXiv preprint arXiv:2502.19643}, 2025.

\bibitem{zhao20233}
L.-W. Zhao, Y.~F. Wu, C.~Wang, and Y.~Guo, ``A 3-d-printed deployable luneburg lens antenna based on the pop-up kirigami sphere,'' \emph{IEEE Transactions on Antennas and Propagation}, vol.~71, no.~8, pp. 6481--6489, 2023.

\bibitem{mitha2021principles}
T.~Mitha and M.~Pour, ``Principles of adaptive element spacing in linear array antennas,'' \emph{Scientific Reports}, vol.~11, no.~1, p. 5584, 2021.

\bibitem{shao20256dma}
X.~Shao and R.~Zhang, ``6dma enhanced wireless network with flexible antenna position and rotation: Opportunities and challenges,'' \emph{IEEE Communications Magazine}, vol.~63, no.~4, pp. 121--128, 2025.

\bibitem{zheng2025tri}
P.~Zheng, Y.~Zhang, T.~Y. Al-Naffouri, M.~J. Hossain, and A.~Chaaban, ``Tri-hybrid multi-user precoding using pattern-reconfigurable antennas: Fundamental models and practical algorithms,'' \emph{arXiv preprint arXiv:2505.08938}, 2025.

\bibitem{lai2023performance}
X.~Lai, T.~Wu, J.~Yao, C.~Pan, M.~Elkashlan, and K.-K. Wong, ``On performance of fluid antenna system using maximum ratio combining,'' \emph{IEEE Communications Letters}, vol.~28, no.~2, pp. 402--406, 2023.

\bibitem{zhu2023modeling}
L.~Zhu, W.~Ma, and R.~Zhang, ``Modeling and performance analysis for movable antenna enabled wireless communications,'' \emph{IEEE Transactions on Wireless Communications}, vol.~23, no.~6, pp. 6234--6250, 2023.

\bibitem{hu2024intelligent}
G.~Hu, Q.~Wu, D.~Xu, K.~Xu, J.~Si, Y.~Cai, and N.~Al-Dhahir, ``Intelligent reflecting surface-aided wireless communication with movable elements,'' \emph{IEEE Wireless Communications Letters}, vol.~13, no.~4, pp. 1173--1177, 2024.

\bibitem{xu2024capacity}
H.~Xu, K.-K. Wong, W.~K. New, F.~R. Ghadi, G.~Zhou, R.~Murch, C.-B. Chae, Y.~Zhu, and S.~Jin, ``Capacity maximization for fas-assisted multiple access channels,'' \emph{IEEE Transactions on Communications}, 2024.

\bibitem{new2023information}
W.~K. New, K.-K. Wong, H.~Xu, K.-F. Tong, and C.-B. Chae, ``An information-theoretic characterization of mimo-fas: Optimization, diversity-multiplexing tradeoff and q-outage capacity,'' \emph{IEEE Trans. Wireless Commun.}, vol.~23, no.~6, pp. 5541--5556, 2023.

\bibitem{new2023fluid}
------, ``Fluid antenna system: New insights on outage probability and diversity gain,'' \emph{IEEE Trans. Wireless Commun.}, vol.~23, no.~1, pp. 128--140, 2023.

\bibitem{wang2019overview}
M.~Wang, F.~Gao, S.~Jin, and H.~Lin, ``An overview of enhanced massive mimo with array signal processing techniques,'' \emph{IEEE Journal of Selected Topics in Signal Processing}, vol.~13, no.~5, pp. 886--901, 2019.

\bibitem{neumann2018learning}
D.~Neumann, T.~Wiese, and W.~Utschick, ``Learning the mmse channel estimator,'' \emph{IEEE Transactions on Signal Processing}, vol.~66, no.~11, pp. 2905--2917, 2018.

\bibitem{cheng2022rethinking}
L.~Cheng, F.~Yin, S.~Theodoridis, S.~Chatzis, and T.-H. Chang, ``Rethinking bayesian learning for data analysis: The art of prior and inference in sparsity-aware modeling,'' \emph{IEEE Signal Processing Magazine}, vol.~39, no.~6, pp. 18--52, 2022.

\bibitem{choi2017compressed}
J.~W. Choi, B.~Shim, Y.~Ding, B.~Rao, and D.~I. Kim, ``Compressed sensing for wireless communications: Useful tips and tricks,'' \emph{IEEE Communications Surveys \& Tutorials}, vol.~19, no.~3, pp. 1527--1550, 2017.

\bibitem{new2024channel}
W.~K. New, K.-K. Wong, H.~Xu, F.~R. Ghadi, R.~Murch, and C.-B. Chae, ``Channel estimation and reconstruction in fluid antenna system: Oversampling is essential,'' \emph{IEEE Transactions on Wireless Communications}, 2024.

\bibitem{wang2023estimation}
R.~Wang, Y.~Chen, Y.~Hou, K.-K. Wong, and X.~Tao, ``Estimation of channel parameters for port selection in millimeter-wave fluid antenna systems,'' in \emph{2023 IEEE/CIC International Conference on Communications in China (ICCC Workshops)}.\hskip 1em plus 0.5em minus 0.4em\relax IEEE, 2023, pp. 1--6.

\bibitem{skouroumounis2022fluid}
C.~Skouroumounis and I.~Krikidis, ``Fluid antenna with linear mmse channel estimation for large-scale cellular networks,'' \emph{IEEE Trans. Commun.}, vol.~71, no.~2, pp. 1112--1125, 2022.

\bibitem{zhang2024successive}
Z.~Zhang, J.~Zhu, L.~Dai, and R.~W. Heath, ``Successive bayesian reconstructor for channel estimation in fluid antenna systems,'' \emph{IEEE Transactions on Wireless Communications}, 2024.

\bibitem{ma2023compressed}
W.~Ma, L.~Zhu, and R.~Zhang, ``Compressed sensing based channel estimation for movable antenna communications,'' \emph{IEEE Commun. Lett.}, vol.~27, no.~10, pp. 2747--2751, 2023.

\bibitem{xiao2024channel}
Z.~Xiao, S.~Cao, L.~Zhu, Y.~Liu, B.~Ning, X.-G. Xia, and R.~Zhang, ``Channel estimation for movable antenna communication systems: A framework based on compressed sensing,'' \emph{IEEE Trans. Wireless Commun.}, 2024.

\bibitem{cao2024channel}
S.~Cao, L.~Zhu, X.~Pi, Z.~Xiao, and B.~Ning, ``Channel estimation for movable antenna communication systems based on compressed sensing,'' in \emph{2024 IEEE Wireless Communications and Networking Conference (WCNC)}.\hskip 1em plus 0.5em minus 0.4em\relax IEEE, 2024, pp. 1--6.

\bibitem{xu2023channel}
H.~Xu, G.~Zhou, K.-K. Wong, W.~K. New, C.~Wang, C.-B. Chae, R.~Murch, S.~Jin, and Y.~Zhang, ``Channel estimation for fas-assisted multiuser mmwave systems,'' \emph{IEEE Commun. Lett.}, vol.~28, no.~3, pp. 632--636, 2023.

\bibitem{zhang2024channel}
R.~Zhang, L.~Cheng, W.~Zhang, X.~Guan, Y.~Cai, W.~Wu, and R.~Zhang, ``Channel estimation for movable-antenna mimo systems via tensor decomposition,'' \emph{IEEE Wireless Commun. Lett.}, 2024.

\bibitem{ji2024correlation}
S.~Ji, C.~Psomas, and J.~Thompson, ``Correlation-based machine learning techniques for channel estimation with fluid antennas,'' in \emph{2024 IEEE International Conference on Acoustics, Speech and Signal Processing (ICASSP)}.\hskip 1em plus 0.5em minus 0.4em\relax IEEE, 2024, pp. 8891--8895.

\bibitem{tang2025accurate}
E.~Tang, W.~Guo, H.~He, S.~Song, J.~Zhang, and K.~B. Letaief, ``Accurate and fast channel estimation for fluid antenna systems with diffusion models,'' \emph{arXiv preprint arXiv:2505.04930}, 2025.

\bibitem{bajwa2008learning}
W.~U. Bajwa, A.~M. Sayeed, and R.~Nowak, ``Learning sparse doubly-selective channels,'' in \emph{2008 46th Annual Allerton Conference on Communication, Control, and Computing}.\hskip 1em plus 0.5em minus 0.4em\relax IEEE, 2008, pp. 575--582.

\bibitem{srivastava2021sparse}
S.~Srivastava, C.~S.~K. Patro, A.~K. Jagannatham, and L.~Hanzo, ``Sparse, group-sparse, and online bayesian learning aided channel estimation for doubly-selective mmwave hybrid mimo ofdm systems,'' \emph{IEEE Transactions on Communications}, vol.~69, no.~9, pp. 5843--5858, 2021.

\bibitem{wei2022off}
Z.~Wei, W.~Yuan, S.~Li, J.~Yuan, and D.~W.~K. Ng, ``Off-grid channel estimation with sparse bayesian learning for otfs systems,'' \emph{IEEE Transactions on Wireless Communications}, vol.~21, no.~9, pp. 7407--7426, 2022.

\bibitem{shen2019channel}
W.~Shen, L.~Dai, J.~An, P.~Fan, and R.~W. Heath, ``Channel estimation for orthogonal time frequency space (otfs) massive mimo,'' \emph{IEEE Transactions on Signal Processing}, vol.~67, no.~16, pp. 4204--4217, 2019.

\bibitem{candes2011compressed}
E.~J. Candes, Y.~C. Eldar, D.~Needell, and P.~Randall, ``Compressed sensing with coherent and redundant dictionaries,'' \emph{Appl. Comput. Harmon. Anal.}, vol.~31, no.~1, pp. 59--73, 2011.

\bibitem{eiwen2010group}
D.~Eiwen, G.~Taub{\"o}ck, F.~Hlawatsch, and H.~G. Feichtinger, ``Group sparsity methods for compressive channel estimation in doubly dispersive multicarrier systems,'' in \emph{2010 IEEE 11th International Workshop on Signal Processing Advances in Wireless Communications (SPAWC)}.\hskip 1em plus 0.5em minus 0.4em\relax IEEE, 2010, pp. 1--5.

\bibitem{taubock2010compressive}
G.~Taubock, F.~Hlawatsch, D.~Eiwen, and H.~Rauhut, ``Compressive estimation of doubly selective channels in multicarrier systems: Leakage effects and sparsity-enhancing processing,'' \emph{IEEE J. Sel. Top. Signal Process.}, vol.~4, no.~2, pp. 255--271, 2010.

\bibitem{eldar2009robust}
Y.~C. Eldar and M.~Mishali, ``Robust recovery of signals from a structured union of subspaces,'' \emph{IEEE Trans. Inf. Theory}, vol.~55, no.~11, pp. 5302--5316, 2009.

\bibitem{eldar2010block}
Y.~C. Eldar, P.~Kuppinger, and H.~Bolcskei, ``Block-sparse signals: Uncertainty relations and efficient recovery,'' \emph{IEEE Trans. Signal Process.}, vol.~58, no.~6, pp. 3042--3054, 2010.

\bibitem{swirszcz2009grouped}
G.~Swirszcz, N.~Abe, and A.~C. Lozano, ``Grouped orthogonal matching pursuit for variable selection and prediction,'' \emph{Advances in Neural Information Processing Systems}, vol.~22, 2009.

\bibitem{dong2024wireless}
X.~Dong, X.~Ren, B.~Lai, R.~Xiong, T.~Mi, and R.~C. Qiu, ``Wireless communications in cavity: A reconfigurable boundary modulation based approach,'' in \emph{ICC 2024-IEEE International Conference on Communications}.\hskip 1em plus 0.5em minus 0.4em\relax IEEE, 2024, pp. 5553--5558.

\bibitem{chai2022port}
Z.~Chai, K.-K. Wong, K.-F. Tong, Y.~Chen, and Y.~Zhang, ``Port selection for fluid antenna systems,'' \emph{IEEE Communications Letters}, vol.~26, no.~5, pp. 1180--1184, 2022.

\bibitem{xiao2024multiuser}
Z.~Xiao, X.~Pi, L.~Zhu, X.-G. Xia, and R.~Zhang, ``Multiuser communications with movable-antenna base station: Joint antenna positioning, receive combining, and power control,'' \emph{IEEE Transactions on Wireless Communications}, 2024.

\bibitem{qiu1999multipath}
R.~C. Qiu and I.-T. Lu, ``Multipath resolving with frequency dependence for wide-band wireless channel modeling,'' \emph{IEEE Trans. Veh. Technol.}, vol.~48, no.~1, pp. 273--285, 1999.

\bibitem{qiu2002study}
R.~C. Qiu, ``A study of the ultra-wideband wireless propagation channel and optimum uwb receiver design,'' \emph{IEEE J. Sel. Areas Commun.}, vol.~20, no.~9, pp. 1628--1637, 2002.

\bibitem{candes2006stable}
E.~J. Candes, J.~K. Romberg, and T.~Tao, ``Stable signal recovery from incomplete and inaccurate measurements,'' \emph{Commun. Pure Appl. Math.}, vol.~59, no.~8, pp. 1207--1223, 2006.

\bibitem{candes2006robust}
E.~J. Cand{\`e}s, J.~Romberg, and T.~Tao, ``Robust uncertainty principles: Exact signal reconstruction from highly incomplete frequency information,'' \emph{IEEE Trans. Inf. Theory}, vol.~52, no.~2, pp. 489--509, 2006.

\bibitem{tropp2004greed}
J.~A. Tropp, ``Greed is good: Algorithmic results for sparse approximation,'' \emph{IEEE Trans. Inf. Theory}, vol.~50, no.~10, pp. 2231--2242, 2004.

\bibitem{donoho2001uncertainty}
D.~L. Donoho, X.~Huo \emph{et~al.}, ``Uncertainty principles and ideal atomic decomposition,'' \emph{IEEE Trans. Inf. Theory}, vol.~47, no.~7, pp. 2845--2862, 2001.

\bibitem{elad2002generalized}
M.~Elad and A.~M. Bruckstein, ``A generalized uncertainty principle and sparse representation in pairs of bases,'' \emph{IEEE Trans. Inf. Theory}, vol.~48, no.~9, pp. 2558--2567, 2002.

\bibitem{wang2020sparse}
X.~Wang and E.~Aboutanios, ``Sparse array design for multiple switched beams using iterative antenna selection method,'' \emph{Digital Signal Process.}, vol. 105, p. 102684, 2020.

\bibitem{gao2017massive}
Y.~Gao, H.~Vinck, and T.~Kaiser, ``Massive mimo antenna selection: Switching architectures, capacity bounds, and optimal antenna selection algorithms,'' \emph{IEEE Trans. Signal Process.}, vol.~66, no.~5, pp. 1346--1360, 2017.

\bibitem{mendoncca2019antenna}
M.~O. Mendon{\c{c}}a, P.~S. Diniz, T.~N. Ferreira, and L.~Lovisolo, ``Antenna selection in massive mimo based on greedy algorithms,'' \emph{IEEE Trans. Wireless Commun.}, vol.~19, no.~3, pp. 1868--1881, 2019.

\bibitem{huang2018wideband}
W.~Huang, Y.~Huang, Y.~Zeng, and L.~Yang, ``Wideband millimeter wave communication with lens antenna array: Joint beamforming and antenna selection with group sparse optimization,'' \emph{IEEE Trans. Wireless Commun.}, vol.~17, no.~10, pp. 6575--6589, 2018.

\bibitem{rasheed2020sparse}
O.~K. Rasheed, G.~Surabhi, and A.~Chockalingam, ``Sparse delay-doppler channel estimation in rapidly time-varying channels for multiuser otfs on the uplink,'' in \emph{2020 IEEE 91st vehicular technology conference (VTC2020-Spring)}.\hskip 1em plus 0.5em minus 0.4em\relax IEEE, 2020, pp. 1--5.

\bibitem{han2016two}
Y.~Han and J.~Lee, ``Two-stage compressed sensing for millimeter wave channel estimation,'' in \emph{2016 IEEE International Symposium on Information Theory (ISIT)}.\hskip 1em plus 0.5em minus 0.4em\relax IEEE, 2016, pp. 860--864.

\bibitem{xiong2024optimal}
R.~Xiong, X.~Dong, T.~Mi, K.~Wan, and R.~C. Qiu, ``Optimal discrete beamforming of ris-aided wireless communications: An inner product maximization approach,'' in \emph{2024 IEEE Wireless Communications and Networking Conference (WCNC)}.\hskip 1em plus 0.5em minus 0.4em\relax IEEE, 2024, pp. 1--6.

\end{thebibliography}

\end{document}